\documentclass[11pt, a4paper]{article}

\usepackage{jheppub}
\author[a]{Xenia de la Ossa,}
\author[a]{Mateo Galdeano}
\affiliation[a]{Mathematical Institute, Oxford University\\Andrew Wiles Building, Woodstock Road\\Oxford OX2 6GG, UK}
\emailAdd{delaossa@maths.ox.ac.uk}
\emailAdd{mateo.galdeano@maths.ox.ac.uk}

\usepackage{cleveref}

\usepackage{physics}
\usepackage{tensor}

\usepackage{multirow}
\usepackage{diagbox}

\newcommand*\Z{\mathbb{Z}}
\newcommand*\R{\mathbb{R}}
\newcommand*\C{\mathbb{C}}
\newcommand*\Sc{\mathbb{S}}

\newcolumntype{C}[1]{>{\centering\let\newline\\\arraybackslash\hspace{0pt}}m{#1}}

\begin{document}

\title{Families of solutions of the heterotic G$_2$ system}

\abstract{We construct new families of solutions of the heterotic G$_2$ system  on squashed homogeneous 3-Sasakian manifolds, that is, using squashed metrics on either the 7-sphere or the Aloff-Wallach space $N_{1,1}$. We obtain AdS$_3$ solutions for all values of the squashing parameter $s$ except for the nearly-parallel G$_2$ value $s=\scriptstyle{1/\sqrt{5}} \,$, for which we don't find any solutions. Along the process, we construct different G$_2$-instanton connections on bundles over these squashed manifolds.
}

\maketitle

\section{Introduction}

The heterotic string enjoys very particular features that are not present in type II string theories. First and foremost, the theory comes with a gauge sector at a perturbative level. This is not the case in type II superstrings, where non-abelian gauge fields appear only when considering the full non-perturbative effects due to the presence of D-branes.
This unfortunately comes at a cost: the presence of the Green-Schwarz anomaly cancellation condition imposes highly non-trivial relations between the gauge bundle, the flux and the geometry at first order in the string parameter $\alpha'$. This poses an additional challenge to the construction of heterotic background compactifications with maximally supersymmetric spacetime.

In this paper we study  heterotic compactifications on 7-dimensional manifolds giving 3-dimensional AdS$_3$ spacetime quantum field theories with minimal supersymmetry $\mathcal{N}=1$. The geometry of these compactifications is a warped product of AdS$_3$ and a 7-dimensional compact manifold together with a gauge bundle with a connection. We call the geometrical structure which preserves supersymmetry a \emph{heterotic} G$_2$ \emph{system}, and solutions of this system satisfy the equations of motion of the theory. These compactifications were first studied in references \cite{Gunaydin:1995ku,Gauntlett:2001ur, Friedrich:2001nh, Friedrich:2001yp, Gauntlett:2002sc, Gauntlett:2003cy, Ivanov:2003nd, Ivanov:2009rh, Kunitomo:2009mx, Lukas:2010mf, Gray:2012md}.

There are various properties that make these systems attractive. For instance, they allow not only Minkowski but also Anti-de Sitter spacetime solutions. This interesting feature is absent in the Hull-Strominger system \cite{Strominger:1986uh, Hull:1986kz} which gives supersymmetric four dimensional theories on Minkowski space when compactifying on 6-dimensional manifolds.

The 7-dimensional compact manifolds relevant to heterotic compactifications must possess an \emph{integrable} G$_2$ \emph{structure} \cite{Gauntlett:2001ur, Friedrich:2001yp} which guarantees the existence of a unique connection with totally antisymmetric torsion \cite{Friedrich:2001nh}.
In addition, the perturbative heterotic string  comes with a gauge bundle with a connection which must be a G$_2$ instanton \cite{Gauntlett:2002sc, Gauntlett:2003cy,Ivanov:2003nd,ReyesCarrion:1998si}. These are higher-dimensional analogues of the 4-dimensional Anti Self-Dual (ASD) instantons \cite{donaldson1990geometry} and they have been gathering attention both in the mathematics and physics communities for a long time, see \cite{Gunaydin:1995ku,Donaldson:1996kp,donaldson2009gauge, Earp:2011dh, SaEarp:2011mp,2011arXiv1109.6609W,2011arXiv1101.0880E,2013arXiv1310.7933E,2015arXiv150501080W,2015arXiv151003836M, Calvo-Andrade:2016fti,2017arXiv170306329H, Lotay:2018gxe, Driscoll:2019gad, DelZotto:2021ydd} for a non-exhaustive list of works on this topic. Thus, the heterotic G$_2$ system constitutes a natural environment for the study of these connections. The heterotic G$_2$ system also requires an anomaly cancellation condition which guarantees that supersymmetric solutions satisfy the equations of motion \cite{Hull:1986kz,Ivanov:2009rh,Martelli:2010jx}.

Finally, the infinitesimal moduli space of heterotic G$_2$ systems was recently studied \cite{delaOssa:2016ivz, Clarke:2016qtg, delaOssa:2017pqy, delaOssa:2017gjq, Fiset:2017auc,Clarke:2020erl}. The anomaly cancellation condition plays a major role here: the G$_2$ structure and the instanton connections on the compact manifold are intertwined with the flux. This requirement is very strong.  Ignoring the anomaly cancellation condition gives an  infinitesimal moduli space which is 
infinite dimensional (except when the compactifying manifold has G$_2$ holonomy). However, for the full heterotic G$_2$ system, it turns out to be finite dimensional. One of our main goals is to construct solutions that are amenable to a deformation theory interpretation.

Unfortunately, not many explicit solutions to the heterotic G$_2$ system with minimal supersymmetry are available in the literature despite the efforts of the community in recent years \cite{Fernandez:2008wla, Nolle:2010nn, Fernandez:2014pfa, Clarke:2020erl, Lotay:2021eog}. Indeed, some of these solutions have been shown \cite{delaOssa:2021cgd} to have more than one supersymmetry. In this paper, we construct new families of solutions that present all the interesting features that we have mentioned: AdS$_3$ spacetime, a collection of G$_2$-instantons on the compact G$_2$-structure manifold, and the possibility of regarding the family as a finite version of the infinitesimal deformations of \cite{delaOssa:2017pqy}.

In the remainder of this introductory section we review the geometry of the  7-dimensional manifolds which allow for the preservation of minimal $\mathcal{N}=1$ supersymmetry in three dimensions. As we have already stressed, 
the 7-dimensional compact manifold must admit an integrable G$_2$-structure. Therefore, we begin with a review of G$_2$-structures in \cref{sec:G2Structures}. In \cref{sec:Heteroticg2Systems} we review the heterotic G${}_2$ system in detail, describing the geometric structures that are required to obtain supersymmetric vacua. In particular we briefly point out how Anti-de Sitter 3-dimensional spacetimes emerge.

In \cref{HomogeneousSasakianSolutions} we focus on specific examples of compact 7-manifolds with an integrable  G$_2$ structure. We will later use them to construct solutions of the heterotic G$_2$ system with AdS$_3$ spacetime. This had only been achieved so far for the solutions in \cite{Lotay:2021eog}.
We review in \cref{3Sasakianmanifolds} the definition and mathematical properties of 3-Sasakian manifolds. These manifolds are naturally equiped with a G$_2$-structure that can be deformed by rescaling the metric along certain SU(2) fibres. This process is known as \emph{squashing} and we call these manifolds with squashed metrics \emph{squashed 3-Sasakian} manifolds. 
In \cref{sec:Homogeneouscase} we specialise to the case of homogenous compact squashed 3-Sasakian manifolds. To be precise, these are the squashed 7-sphere and the squashed Aloff-Wallach space. Squashed 7-spheres first appeared in the physics literature in the context of compactifications of 11-dimensional supergravity down to 4 dimensions, see for example \cite{Awada:1982pk, Duff:1983gh, Duff:1986hr}. 

Instanton connections are the centre of attention of \cref{sec:InstantonConnections}. In particular, we describe several G$_2$-instantons on bundles over squashed 3-Sasakian manifolds---with special emphasis on the tangent bundle and the case of homogeneous manifolds.
In \cref{sec:CanonicalConnection} we review the construction of the canonical connection \cite{kobayashi1963foundations,Harland:2010ix}  and we check it is a G$_2$-instanton. In \cref{sec:ClarkeOliveiraConnection} we extend a construction of a G$_2$-instanton by Clarke and Oliveira \cite{2019arXiv190305526C} for all squashed metrics and different representations.  
Lastly, in \cref{sec:tangentbundleinstantons} we describe explicitly a one-parameter family of instantons on the tangent bundle for both the squashed 7-sphere and the squashed Aloff-Wallach space.

We then proceed in \cref{sec:HeteroticBI} to set up the Bianchi identity for the anomaly cancellation condition. This identity is a rather complicated relation involving all the theory's degrees of freedom, that is: the torsion of a G$_2$ compatible connection the manifold, the curvature of the instanton connection on the gauge vector bundle, and the curvature of the instanton connection on the tangent space. The relevant curvature terms can be found in \cref{sec:tracecanonicalconnection,sec:traceclarkeoliveira,sec:tracetangentbundle} for each of the instanton connections we consider.

Finally, we put all the previous data together to construct explicit solutions of the heterotic G$_2$ system. As anticipated, the highly non-trivial heterotic Bianchi identity significantly constrains the potential solutions. For that reason, we can only find solutions for particular combinations of instantons. This is explained in full detail in \cref{sec:NewSolutions}. We have included tables detailing the ranges of the solutions obtained as well as figures illustrating their behaviour.

We comment on our results and point out several possible future directions in \cref{sec:conclusions}. We have also included various appendices collecting some of the relevant quantities for the computations we have performed.

\subsection{G${}_2$ structures}
\label{sec:G2Structures}

A G$_2$ \emph{structure}\footnote{For more detailed accounts on G$_2$ structures, see \cite{Joyce2000} or \cite{2003math......5124B}} on a manifold $Y$ is a reduction of the structure group of the tangent bundle of $Y$ to the group G$_2$. This condition is equivalent to the existence of a non-degenerate positive three-form $\varphi$ on $Y$ that we call the \emph{associative} three-form. Locally, this form can be written as
\begin{equation}
\label{varphi0}
\varphi=e^{123}+e^{145}+e^{167}+e^{246}-e^{257}-e^{347}-e^{356} \, .
\end{equation}
where $\lbrace e^1,\dots,e^7\rbrace$ form an orthonormal basis of one-forms and we are writing $e^{ij}=e^i\wedge e^j \, $. The four-form $\psi=*\varphi$ is called the \emph{coassociative} four-form and it locally takes the form
\begin{equation}
\psi=e^{1357}+e^{2345}+e^{2367}+e^{4567}-e^{1247}-e^{1256}-e^{1346} \, .
\end{equation}

If $Y$ admits a G$_2$ structure, we can decompose all tensors in terms of representations of the group G${}_2$. For p-forms we have
\begin{align}
\label{splittingofformseq}
\begin{split}
\Lambda^0&=\Lambda^0_1 \, ,\\
\Lambda^1&=\Lambda^1_7 \, ,\\
\Lambda^2&=\Lambda^2_7\oplus\Lambda^2_{14} \, ,\\
\Lambda^3&=\Lambda^3_1\oplus\Lambda^3_7\oplus\Lambda^3_{27} \, .
\end{split}
\end{align}
where $\Lambda^k$ denotes the space of k-forms on $Y$ and $\Lambda^k_p$ denotes the subspace of $\Lambda^k$ consisting of k-forms transforming in the $p$-dimensional irreducible representation of G$_2$. The decomposition for higher degrees follows from Hodge duality. The associative and coassociative forms can be used to build projection operators to these subspaces, see for example \cite{Grigorian:2011ap}. To illustrate this, note that $\Lambda^2_{14} \, $, which corresponds to the two-forms contained in the Lie algebra of G${}_2$, can be described as
\begin{equation}
\label{splitofg2liealgebra}
\Lambda^2_{14}=\lbrace\beta\in\Lambda^2 : \beta\lrcorner \, \varphi=0\rbrace=\lbrace\beta\in\Lambda^2 : \beta\wedge\psi=0\rbrace \, .
\end{equation}

We say that a manifold $Y$ has G$_2$ \emph{ holonomy} if it has a G$_2$ structure $\varphi$ such that which is covariantly constant with respect to the Levi-Civita connection.
This condition is equivalent to
\begin{equation}
    \dd\varphi=0 \, , \qquad \dd\psi=0 \, ,
\end{equation}
and we then say that the G$_2$ structure is \emph{torsion-free}. For a general G$_2$ structure, the exterior derivatives of the associative and coassociative forms are non-zero and its decomposition in G$_2$ representations gives the \emph{torsion classes} of the G$_2$ structure:
\begin{align}
\label{torsionclassequation}
\begin{split}
\dd\varphi=&\tau_0 \, \psi+3\tau_1\wedge\varphi+*\tau_3 \, ,\\
\dd\psi=&4\tau_1\wedge\psi+*\tau_2 \, .
\end{split}
\end{align}
Here the torsion classes $\tau_k$ are $k$-forms with $\tau_3\in\Lambda^3_{27}$ and $\tau_2\in\Lambda^2_{14}$ \, . We are interested in G$_2$ structures such that $\tau_2=0 \, $, which are called \emph{integrable}. Given an integrable G$_2$ structure there exists a unique metric connection compatible with the G$_2$ structure with totally antisymmetric torsion \cite{2003math......5124B} given by the three-form
\begin{equation}
\label{eq:torsiong2}
T(\varphi)=\frac{1}{6}\tau_0 \, \varphi-\tau_1\lrcorner \, \psi-\tau_3 \, .
\end{equation}
We call this the \emph{torsion} of the G$_2$ structure.

\subsection{Heterotic G$_2$ systems}
\label{sec:Heteroticg2Systems}

We now introduce heterotic G${}_2$ systems following \cite{delaOssa:2017pqy}. See also \cite{Gauntlett:2001ur, Friedrich:2001nh, Friedrich:2001yp, Gauntlett:2002sc, Gauntlett:2003cy, Ivanov:2003nd, Ivanov:2009rh}. These describe $\mathcal{N}=1$ supersymmetric vacuum solutions of heterotic string theory on a manifold of the form $M_{3}\times Y$, where $M_{3}$ is a maximally symmetric 3-dimensional Lorentzian space (the \emph{spacetime}), $Y$ is a compact 7-dimensional manifold and the metric is a warped product. 

A \emph{heterotic G${}_2$ system} is given by a quadruple $[(Y,\varphi),(V,A),(TY,\Theta),H]$ satisfying the following properties:
\begin{itemize}
\item $Y$ is a 7-dimensional manifold and $\varphi$ is a three-form on $Y$ defining an integrable G$_2$ structure on $Y$. We call $\psi=*\varphi \, $.
\item $V$ is a vector bundle on $Y$ with a connection $A$ that is a G$_2$-instanton, so that its curvature $F_A$ satisfies $F_A\wedge \psi=0 \, $.
\item $TY$ is the tangent bundle of $Y$ and $\Theta$ is a connection on $TY$ which is a G$_2$-instanton, so that its curvature $R_\Theta$ satisfies $R_\Theta\wedge \psi=0 \, $.\footnote{We denote the curvature of a gauge connection $A$ by $F_A$ and the curvature of a metric connection $\Theta$ by $R_\Theta \, $. We omit the subscript only if it is clear the connection we are making reference to.}
\item $H$ is a three-form on $Y$ defined by the formula
\begin{equation}
\label{eq:anomalycancellation}
    H=\dd B+\frac{\alpha'}{4}\left(\mathcal{CS}(A)-\mathcal{CS}(\Theta)\right) \, ,
\end{equation}
where $\mathcal{CS}$ denotes the Chern-Simons form of the corresponding connection, $B$ is the antisymmetric two-form and $\alpha'>0$ is the string parameter. In addition, $H$ is constrained to satisfy
\begin{equation}
\label{eq:fluxistorsion}
    H=T(\varphi)=\frac{1}{6}\,\tau_0\,\varphi-\tau_1\lrcorner\psi-\tau_3 \, .
\end{equation}
where $T(\varphi)$ denotes the torsion three-form of the G$_2$ structure as defined in \eqref{eq:torsiong2}.
\end{itemize}

\smallskip

These conditions guarantee that we have a solution of the Killing spinor equations, which result from preserving $\mathcal{N}=1$ supersymmetry,  and together with the anomaly cancellation condition, guarantee a solution of the equations of motion of the 10-dimensional supergravity action. 

Taking the exterior derivative of \eqref{eq:anomalycancellation} we obtain the \emph{heterotic Bianchi identity}
\begin{equation}
\label{eq:heteroticBianchiidentity}
\dd \mathcal{H}=\frac{\alpha'}{4}(\tr F_A\wedge F_A -\tr R_\Theta\wedge R_\Theta) \, ,
\end{equation}
and finding a solution of \eqref{eq:heteroticBianchiidentity} automatically ensures that a solution to the anomaly cancellation condition \eqref{eq:anomalycancellation} exists.

As pointed out in \cite{delaOssa:2019jsx}, the cosmological constant of the 3-dimensional spacetime is related to the external component of the flux $h$, which is determined by the torsion classes of the compact manifold: $h=\frac{1}{3}\tau_0$. Therefore,
\begin{equation}
\label{eq:tau0givescurvature}
\Lambda\sim -h^2 \implies \Lambda\sim -\tau_0^2 \, ,
\end{equation}
and we obtain that for $\tau_0\neq 0$ the non-compact spacetime is AdS$_3$. For a Minkowski spacetime, $\tau_0=0$ and, due to the dilaton condition $\tau_1=\frac{1}{2}\dd\phi \, $, the G$_2$ structure is cocalibrated of pure type as observed in \cite{Friedrich:2001yp}.

\bigskip

\section{Squashed 3-Sasakian manifolds}
\label{HomogeneousSasakianSolutions}

\subsection{General aspects, squashing and G$_2$ structures}
\label{3Sasakianmanifolds}

In this section we introduce 3-Sasakian manifolds, which are one of the main elements in our solutions of the heterotic G${}_2$ system. A more detailed account of 3-Sasakian manifolds can be found in \cite{Boyer:1998sf} and \cite{Boyer:2007nr}. See also \cite{2010JGP....60..326A} for the 7-dimensional case.

Let $(Y,g)$ be a Riemannian manifold of dimension $n$, where $n=4k+3$ for $k\geq 1 \, $. We say $(Y,g)$ is \emph{3-Sasakian} if its metric cone $(C(Y),\Bar{g})=(\mathbb{R}_+\times Y,\dd r^2+r^2 g)$
is a hyperk\"{a}hler manifold. We are interested in the 7-dimensional case so from now on we fix $n=7 \, $.

Every 3-Sasakian manifold $Y$ has a triple of orthonormal Killing vector fields $(\xi_1, \xi_2, \xi_3)$  satisfying the relation $[\xi_i,\xi_j]=2\epsilon\indices{_{ij}^k}\xi_k \, $, where $\epsilon_{ijk}$ is the Levi-Civita symbol.  It then follows that these Killing vector fields form an integrable distribution and define a 3-dimensional foliation of $Y$. Moreover, it turns out that the space of leaves of this foliation is a compact orbifold and we can think of $Y$ as the total space of a bundle over an orbifold\footnote{The simplest example of a 3-Sasakian manifold is the 7-sphere $\Sc^7$, which is the total space of an SU(2)-bundle over the 4-sphere as described by the Hopf fibration $\Sc^3\longrightarrow\Sc^7\longrightarrow\Sc^4$.}.

We can locally complete $(\xi_1, \xi_2, \xi_3)$ to an orthonormal basis $\lbrace \xi_1, \dots , \xi_7\rbrace$ of the 3-Sasakian manifold $Y$ and work with the dual basis of one-forms $\lbrace \xi^1, \dots , \xi^7\rbrace$. We define for later convenience:
\begin{equation}
\label{defofomegas}
\omega^1=\xi^4\wedge\xi^5+\xi^6\wedge\xi^7 \, , \qquad
\omega^2=\xi^4\wedge\xi^6-\xi^5\wedge\xi^7 \, , \qquad
\omega^3=-\xi^4\wedge\xi^7-\xi^5\wedge\xi^6 \, .
\end{equation}
and it can be shown that the following formulas hold:
\begin{align}
\label{eq:derivativesofetas}
\dd \xi^i&=2 \, \omega^i-\epsilon\indices{^i_{jk}}\,\xi^j\wedge \xi^k \, , \\
\label{eq:derivativesofomegas}
\dd \omega^i&=-2 \, \epsilon\indices{^i_{jk}}\,\xi^j\wedge \omega^k \, ,
\end{align}
where $i,j,k\in\{1,2,3\}$.

We can deform the metric of $Y$ away from the 3-Sasakian metric by rescaling the metric along the fibres while keeping the base orbifold metric fixed. This process is known as \emph{squashing} and we obtain a one-parameter family of metrics
\begin{equation}
\label{eq:squashedmetric}
ds^2=\sum_{i=1}^3 s^2\, \xi^i\otimes \xi^i+\sum_{\alpha=4}^7 \xi^\alpha\otimes \xi^\alpha \, ,
\end{equation}
where $s>0$ is the \emph{squashing parameter} and we recover the original metric for $s=1 \, $. We call these manifolds \emph{squashed 3-Sasakian manifolds}. It will be convenient to define an orthonormal coframe $\{\eta^1,\dots,\eta^7\}$ for each value of $s$
\begin{equation}
\label{eq:coframe}
\eta^i=s \, \xi^i \qquad \text{for} \ i=1,2,3 \, ; \qquad \qquad
\eta^\alpha=\xi^\alpha \qquad \text{for} \ \alpha=4,5,6,7.
\end{equation}
The two-forms $\omega^i$ from \eqref{defofomegas} have an analogous expression in this basis, and the formulas \eqref{eq:derivativesofetas} and \eqref{eq:derivativesofomegas} now take the form
\begin{align}
\label{eq:derivativesofetasimproved}
\dd \eta^i&=2 \, s\,\omega^i-\frac{1}{s}\epsilon\indices{^i_{jk}}\,\eta^j\wedge \eta^k \, , \\
\label{eq:derivativesofomegasimproved}
\dd \omega^i&=-\frac{2}{s}\,\epsilon\indices{^i_{jk}}\,\eta^j\wedge \omega^k \, ,
\end{align}
making manifest that the 3-Sasakian structure is lost by the squashing procedure.

We can define a G$_2$-structure on $Y$ for each value of the squashing parameter. The associative three-form is given by\footnote{For $s=1$ this is called the \emph{canonical} G$_2$-structure of the 3-Sasakian manifold in \cite{2010JGP....60..326A}.}
\begin{equation}
\label{eq:threeformforsquash}
\varphi_s=\eta^{123}+\eta^1\wedge\omega^1+\eta^2\wedge\omega^2+\eta^3\wedge\omega^3\, ,
\end{equation}
where $\eta^{\mu\nu\rho}=\eta^\mu\wedge\eta^\nu\wedge\eta^\rho$. It is important to remark that the structure depends on the parameter $s$. This can be seen explicitly writing \eqref{eq:threeformforsquash} in terms of the $\{\xi^\mu\}$ basis:
\begin{equation}
\label{threeforminotherbasis}
\varphi_s=s^3\,\xi^{123}+s\,\xi^1\wedge\omega^1+s\,\xi^2\wedge\omega^2+s\,\xi^3\wedge\omega^3 \, .
\end{equation}
Therefore we are defining a \emph{one-parameter family} of G$_2$-structures that change with the metric as the squashing parameter $s$ varies. Following \cite{FRIEDRICH1997259}, we rewrite $\varphi_s=F_1+F_2$ with
\begin{equation}
F_1=\eta^{123} \, , \qquad
F_2=\eta^1\wedge\omega^1+\eta^2\wedge\omega^2+\eta^3\wedge\omega^3 \, ,
\end{equation}
we can then compute
\begin{equation}
\label{eq:dualF1andF2}
*_s F_1=\frac{1}{6}\sum_{i=1}^3\omega^i\wedge\omega^i \, , \qquad
*_s F_2=\frac{1}{2}\epsilon_{ijk}\,\eta^i\wedge \eta^j\wedge \omega^k \, ,
\end{equation}
and
\begin{equation}
\label{eq:exterdersF1andF2}
\dd F_1=2\,s*_sF_2 \, , \qquad
\dd F_2=12\,s*_sF_1+2\,\frac{1}{s}*_sF_2 \, ,
\end{equation}
where $*_s$ is the Hodge star with respect to the metric associated with the G$_2$-structure $\varphi_s \, $. This determines the coassociative four-form
\begin{equation}
\label{eq:fourformsquashed}
\psi_s=*_s\varphi_s=*_s F_1+*_s F_2 \, .
\end{equation}

We can then compute the exterior derivative of the G$_2$-forms
\begin{align}
\label{eq:extderG2squashed}
\begin{split}
\dd\varphi_s=&\frac{12}{7}\left(2 \, s+\frac{1}{s}\right)\psi_s+\left(10 \, s-\frac{2}{s}\right)\left( *_s F_1-\frac{1}{7}\psi_s \right),\\
\dd\psi_s=&0 \, ,
\end{split}
\end{align}
and the torsion classes of the G$_2$-structure can be extracted from these expressions as in equation \eqref{torsionclassequation}.
\begin{align}
\label{eq:tau0}
\tau_0(\varphi_s)&=\frac{12}{7}\left(2 \, s+\frac{1}{s}\right),\\
\label{eq:tau1}
\tau_1(\varphi_s)&=0 \, ,\\
\label{eq:tau2}
\tau_2(\varphi_s)&=0 \, ,\\
\tau_3(\varphi_s)&=\left(10 \, s-\frac{2}{s}\right)\left( F_1-\frac{1}{7}\varphi_s \right).
\label{eq:tau3}
\end{align}
For all values of the squashing parameter the G$_2$-structure is coclosed. In particular it is always integrable, $\tau_2=0 \, $, and thus can be used to construct solutions of the heterotic G${}_2$ system. Since $\tau_1$ vanishes, all these solutions will have constant dilaton. Note as well that $\tau_3$ vanishes if and only if $s=\frac{1}/{\sqrt{5}} \, $, in this case the only nonzero torsion class is $\tau_0$ and we say that the G$_2$-structure is \emph{nearly parallel}.\footnote{We later apply these results to the 7-sphere and it is important to remark that the canonical 3-Sasakian G$_2$-structure we obtain from \eqref{eq:threeformforsquash} when $s=1$ is not nearly-parallel. It is therefore different from the \emph{standard} G$_2$-structure of the round 7-sphere, which is known to be nearly-parallel.}

Since the torsion class $\tau_0$ is nonzero for all values of the squashing parameter $s$, heterotic solutions constructed using these manifolds only give rise to AdS$_3$ spacetimes. This is interesting as the only solutions of this kind available in the literature so far are those of \cite{Lotay:2021eog}.

From \eqref{eq:torsiong2} we find for each value of the parameter $s$ the unique connection which is metric, compatible with the G$_2$-structure and has totally antisymmetric torsion
\begin{equation}
\label{eq:uniqueantitorsion}
T(\varphi_s)=2 \, s \, \varphi_s+\left(\frac{2}{s}-10 \, s\right) F_1 \, ,
\end{equation}
in agreement with \cite{FRIEDRICH2007632}.

\subsection{Homogeneous 3-Sasakian manifolds}
\label{sec:Homogeneouscase}

We turn our attention to \emph{homogeneous} 3-Sasakian manifolds. These are described as coset spaces $G/H$ where $G$ is a Lie group and $H$ is a closed subgroup of $G$. Coset spaces are described in further detail in \cite{kobayashi1963foundations, Kapetanakis:1992hf, Harland:2010ix}. Homogeneous 3-Sasakian manifolds are fully classified, see \cite{Boyer:1998sf}, and in 7 dimensions we only have the 7-sphere $\Sc^7=\text{Sp}(2)/\text{Sp}(1)$ and the squashed Aloff-Wallach space $N_{1,1}=\text{SU}(3)/\text{U}(1)_{1,1} \, $. These are the only regular 3-Sasakian manifolds in dimension 7 \cite{cmp/1104201507} and we present them in sections \ref{sec:The7sphere} and \ref{sec:AloffWallach}.

The coset structure of these manifolds can be used to describe an orthonormal coframe realising the 3-Sasakian structure \eqref{eq:derivativesofetas} and \eqref{eq:derivativesofomegas}, as we now explain. Recall that a Lie group $G$ acts on itself by left translations and  left-invariant vector fields can be identified with elements of the Lie algebra $\mathfrak{g}$, so that left-invariant one-forms are identified with elements of the dual $\mathfrak{g}^*$.

Let $\lbrace I_1,\dots,I_{\rank(\mathfrak{g})}\rbrace$ be a basis of $\mathfrak{g}$ and let $\lbrace e^1,\dots,e^{\rank(\mathfrak{g})}\rbrace$ be the dual basis. The \emph{Maurer-Cartan form} is defined as the unique $\mathfrak{g}$-valued one-form that acts as the identity on the elements of $\mathfrak{g}$ and it's written in a basis as
\begin{equation}
\label{eq:maurercartanform}
\theta=\sum_{\alpha=1}^{\rank(\mathfrak{g})} I_\alpha\otimes e^\alpha \, .
\end{equation}
We can regard the Maurer-Cartan form as a metric on $G$, so the $\mathfrak{g}^*$ basis $\lbrace e^\alpha \rbrace$ is describing an orthonormal coframe of $G$. Moreover, it can be shown from the Maurer-Cartan form that the coframe must satisfy the structure equations
\begin{equation}
\label{eq:generalstructureequation}
\dd e^\alpha=-\frac{1}{2}f^\alpha_{\beta\gamma}\, e^\beta\wedge e^\gamma \, ,
\end{equation}
with $\alpha,\beta,\gamma=1,\dots,\rank(\mathfrak{g}) \, $, and where $f^\alpha_{\beta\gamma}$ are the structure constants of the Lie group $G$.

We are interested in describing a coframe of the coset space $G/H$, to this end it is convenient to consider $G$ as a principal $H$-bundle over $G/H$. Let $\mathfrak{h}$ be the Lie algebra of $H$ and assume the coset is \emph{reductive}, that means there exists a subspace $\mathfrak{m}$ of $\mathfrak{g}$ such that $\mathfrak{g}=\mathfrak{h}\oplus\mathfrak{m}$ and $\mathfrak{m}$ is invariant under the adjoint action of $H$, $[\mathfrak{h},\mathfrak{m}]\subset\mathfrak{m} \, $. Under this assumption $\mathfrak{m}$ is a Lie subalgebra that can be identified with the tangent space of the coset space, see \cite{kobayashi1963foundations}.

We can split the basis of $\mathfrak{g}$ in generators of $\mathfrak{m}$, $\lbrace I_1 \, ,\dots,I_{\rank(\mathfrak{m})}\rbrace$, and generators of $\mathfrak{h}$, $\lbrace I_{\rank(\mathfrak{m})+1} \, ,\dots,I_{\rank(\mathfrak{g})}\rbrace$. The commutation relation can then be written as
\begin{equation}
\label{eq:structureconstants}
[I_b,I_c]=f^a_{bc}\, I_a \, , \qquad [I_b,I_\nu]=f^\mu_{b\nu}\, I_\mu \, , \qquad [I_\nu,I_\rho]=f^a_{\mu\nu}\, I_a+f^\mu_{\nu\rho}\,I_\mu \, ,
\end{equation}
where $\mu,\nu,\rho=1 \, ,\dots,\rank(\mathfrak{m})$ and $a,b,c=\rank(\mathfrak{m})+1 \, ,\dots,\rank(\mathfrak{g})$. The structure equations \eqref{eq:generalstructureequation} can be rewritten as
\begin{equation}
    \label{eq:forGstruceq}
\dd e^\mu=-f^\mu_{a\nu}\, e^a\wedge e^\nu-\frac{1}{2}f^\mu_{\nu\rho}\, e^\nu\wedge e^\rho \, , \qquad
\dd e^a=-\frac{1}{2}f^a_{\nu\rho}\, e^\nu\wedge e^\rho-\frac{1}{2}f^a_{bc}\, e^b\wedge e^c \, ,
\end{equation}
where again $\mu,\nu,\rho=1 \, ,\dots,\rank(\mathfrak{m})$ and $a,b,c=\rank(\mathfrak{m})+1 \, ,\dots,\rank(\mathfrak{g})$.

Take a local patch $U\subset G/H$ and a local section $L$ of the principal bundle $\pi :G\longrightarrow G/H$, that is, a map $L: U\longrightarrow G$ such that $\pi\circ L$ is the identity. We use $L$ to pull back the one-forms $\lbrace e^\alpha\rbrace$ to the coset space, $\lbrace L^*e^\alpha=\xi^\alpha\rbrace$. In particular the one-forms $\lbrace e^\mu\rbrace$ with $\mu=1 \, ,\dots,\rank(\mathfrak{m})$ are pulled back to a coframe $\{\xi^{\mu}\}$ of $G/H$\footnote{The pulled-back one-forms $\xi^a$ with $a=\rank(\mathfrak{m})+1,\dots,\rank(\mathfrak{g})$ can be rewritten in terms of the coframe as $\xi^a=c^a_{\mu}\xi^{\mu}$ for some functions $c^a_{\mu}$. Nevertheless, it is convenient for computations to work with the forms $\xi^a$ directly and we will continue to do so in the rest of the paper.}. The pullback of the Maurer-Cartan form defines a metric for $G/H$ with respect to which $\{\xi^{\mu}\}$ is orthonormal. Note as well that the structure equations are still satisfied by the pulled-back forms
\begin{equation}
\label{eq:struceq}
\dd \xi^\mu=-f^\mu_{a\nu}\,\xi^a\wedge \xi^\nu-\frac{1}{2}f^\mu_{\nu\rho}\, \xi^\nu\wedge \xi^\rho \, , \qquad \dd \xi^a=-\frac{1}{2}f^a_{\nu\rho}\, \xi^\nu\wedge \xi^\rho-\frac{1}{2}f^a_{bc}\, \xi^b\wedge \xi^c \, .
\end{equation}

In the case of 7-dimensional homogeneous 3-Sasakian manifolds the coframe $\lbrace \xi^1 \, , \dots , \xi^7\rbrace$ can be chosen so that equations \eqref{eq:derivativesofetas} and \eqref{eq:derivativesofomegas} are satisfied, making the 3-Sasakian structure explicit. Furthermore, we can consider a squashing of the metric as in \eqref{eq:squashedmetric}: define an orthonormal coframe as we did in \eqref{eq:coframe} and rename $\eta^a=\xi^a$ for $a=8 \, ,\dots,\rank(\mathfrak{g})$, then \eqref{eq:struceq} can be rewritten as
\begin{align}
\label{eq:squashedstruceq1}
\dd \eta^i&=-f^i_{aj}\,\eta^a\wedge \eta^j-s \, f^i_{a\alpha}\,\eta^a\wedge \eta^\alpha-f^i_{j\alpha}\, \eta^j\wedge \eta^\alpha-\frac{1}{2 \, s}f^i_{jk}\, \eta^j\wedge \eta^k-\frac{s}{2}f^i_{\alpha\beta}\, \eta^\alpha\wedge \eta^\beta \, , \\
\label{eq:squashedstruceq2}
\dd \eta^\alpha&=- \, \frac{1}{s}f^\alpha_{ai}\,\eta^a\wedge \eta^i-f^\alpha_{a\beta}\,\eta^a\wedge \eta^\beta-\frac{1}{s}f^\alpha_{j\beta}\, \eta^j\wedge \eta^\beta-\frac{1}{2 \, s^2}f^\alpha_{ij}\, \eta^i\wedge \eta^j-\frac{1}{2}f^\alpha_{\beta\gamma}\, \eta^\beta\wedge \eta^\gamma \, , \\
\label{eq:squashedstruceq3}
\dd \eta^a&=- \, \frac{1}{s}f^a_{i\alpha}\, \eta^i\wedge \eta^\alpha-\frac{1}{2 \, s^2}f^a_{ij}\, \eta^i\wedge \eta^j-\frac{1}{2}f^a_{\alpha\beta}\, \eta^\alpha\wedge \eta^\beta-\frac{1}{2}f^a_{bc}\, \eta^b\wedge \eta^c \, ,
\end{align}
where $i,j,k\in\lbrace 1,2,3\rbrace$, $\alpha,\beta,\gamma\in\lbrace 4 \, ,\dots,7\rbrace$, $a,b,c\in\lbrace 8 \, ,\dots,\rank(\mathfrak{g})\rbrace$ and $s$ is the squashing parameter. We now particularize this result to the two manifolds we are interested in.

\subsubsection{The squashed 7-sphere}
\label{sec:The7sphere}

The 7-sphere is a homogeneous 3-Sasakian manifold given by the coset
\begin{equation}
\Sc^7=\text{Sp}(2)/\text{Sp}(1) \, ,
\end{equation}
where Sp($n$) denotes the quaternionic unitary groups \footnote{For $n=1,2$ we have alternative characterizations due to some accidental isomorphisms: $\text{Sp}(1)\simeq\text{SU}(2)$ and $\text{Sp}(2)\simeq\text{Spin}(5)$.}
\begin{equation}
\text{Sp}(n)=\{M\in \mathcal{M}_{n\times n}(\mathbb{H}) \text{ such that } MM^\dagger=1\}.
\end{equation}
There are two diagonal $\text{Sp}(1)\simeq\text{SU}(2)$ subgroups inside Sp(2). Quotienting one of them gives the 7-sphere, and the quotient map Sp(2)$\longrightarrow\Sc^7$ describes a principal SU(2)-bundle over $\Sc^7$. If we further quotient $\Sc^7$ by the leftover Sp(1), the resulting coset manifold is the 4-sphere and we recover the Hopf fibration $\Sc^3\longrightarrow\Sc^7\longrightarrow\Sc^4$, which makes the 3-Sasakian structure manifest.

As explained above for the general case, we need to specify a convenient basis of the Lie algebra of Sp(2) in order to describe a coframe of the 7-sphere. See also \cite{Geipel:2017tmp} for an alternative equivalent method to obtain a local section using the Hopf fibration. Our conventions can be found in \cref{sec:Sp2struceq}. The coframe structure equations are obtained by substituting the structure constants in \eqref{eq:squashedstruceq1}, \eqref{eq:squashedstruceq2} and \eqref{eq:squashedstruceq3}, obtaining
\begin{align}
\label{s7structureequations}
\begin{split}
\dd\eta^1&= - \, \frac{2}{s}\eta^2\wedge\eta^3 +2 \, s \, \eta^4\wedge\eta^5 +2 \, s \, \eta^6\wedge\eta^7 \, , \\
\dd\eta^2&= - \, \frac{2}{s}\eta^3\wedge\eta^1 +2 \, s \, \eta^4\wedge\eta^6 -2 \, s \, \eta^5\wedge\eta^7 \, , \\
\dd\eta^3&= - \, \frac{2}{s}\eta^1\wedge\eta^2 -2 \, s \, \eta^4\wedge\eta^7 -2 \, s \, \eta^5\wedge\eta^6 \, , \\
\dd\eta^4&= - \, \frac{1}{s}\eta^1\wedge\eta^5 -\frac{1}{s}\eta^2\wedge\eta^6 +\frac{1}{s}\eta^3\wedge\eta^7 -\eta^5\wedge\eta^ 8-\eta^6\wedge\eta^9 +\eta^7\wedge\eta^{10} \, , \\
\dd\eta^5&= + \, \frac{1}{s}\eta^1\wedge\eta^4 +\frac{1}{s}\eta^2\wedge\eta^7 +\frac{1}{s}\eta^3\wedge\eta^6 +\eta^4\wedge\eta^8 -\eta^7\wedge\eta^9 -\eta^6\wedge\eta^{10} \, , \\
\dd\eta^6&= - \, \frac{1}{s}\eta^1\wedge\eta^7 +\frac{1}{s}\eta^2\wedge\eta^4 -\frac{1}{s}\eta^3\wedge\eta^5 +\eta^7\wedge\eta^8 +\eta^4\wedge\eta^9 +\eta^5\wedge\eta^{10} \, , \\
\dd\eta^7&= + \, \frac{1}{s}\eta^1\wedge\eta^6 -\frac{1}{s}\eta^2\wedge\eta^5 -\frac{1}{s}\eta^3\wedge\eta^4 -\eta^6\wedge\eta^8 +\eta^5\wedge\eta^9 -\eta^4\wedge\eta^{10} \, , \\
\dd\eta^8&= -2 \, \eta^9\wedge\eta^{10} -2 \, \eta^4\wedge\eta^5 +2 \, \eta^6\wedge\eta^7 \, , \\
\dd\eta^9&= -2 \, \eta^{10}\wedge\eta^8 -2 \, \eta^4\wedge\eta^6 -2 \, \eta^5\wedge \eta^7 \, , \\
\dd \eta^{10}&= -2 \, \eta^8\wedge\eta^9 +2 \, \eta^4\wedge\eta^7 -2 \, \eta^5\wedge\eta^6 \, .
\end{split}
\end{align}
It can be checked the structure equations satisfy \eqref{eq:derivativesofetasimproved} and \eqref{eq:derivativesofomegasimproved} so the coframe describes the squashed 3-Sasakian structure of the 7-sphere.

\subsubsection{The squashed Aloff-Wallach space}
\label{sec:AloffWallach}

The Aloff-Wallach spaces were first described in \cite{bams/1183536240}, we introduce them following \cite{Ball:2016xte}. Consider the matrix group SU(3), let $k,l\in\Z$ and let U(1)${}_{k,l}$ be the circle subgroup of SU(3) whose elements are of the form:
\begin{equation}
\begin{pmatrix}
e^{ik\theta} &0 &0 \\
0 &e^{il\theta} &0 \\
0 &0 &e^{im\theta}
\end{pmatrix},
\end{equation}
where $k+l+m=0$ and $\theta\in\R$. The \emph{Aloff-Wallach space} $N_{k,l}$ is the coset
\begin{equation}
N_{k,l}=\text{SU}(3)/\text{U}(1)_{k,l} \, .
\end{equation}
The only 3-Sasakian Aloff-Wallach space is $N_{1,1}$ and this is the only case that we will study in this paper.
Our choice of coframe for SU(3) differs slightly from the ones in \cite{Ball:2016xte} and \cite{Geipel:2016hpk} and can be found in \cref{SU3struceq}. After squashing we obtain the following structure equations
\begin{align}
\label{aloffwallachstructureequations}
\begin{split}
\dd\eta^1&= - \, \frac{2}{s}\eta^2\wedge\eta^3 +2 \, s \, \eta^4\wedge\eta^5 +2 \, s \, \eta^6\wedge\eta^7 \, , \\
\dd\eta^2&= - \, \frac{2}{s}\eta^3\wedge\eta^1 +2 \, s \, \eta^4\wedge\eta^6 -2 \, s \, \eta^5\wedge\eta^7 \, , \\
\dd\eta^3&= - \, \frac{2}{s}\eta^1\wedge\eta^2 -2 \, s \, \eta^4\wedge\eta^7 -2 \, s \, \eta^5\wedge\eta^6 \, , \\
\dd\eta^4&= - \, \frac{1}{s}\eta^1\wedge\eta^5 -\frac{1}{s}\eta^2\wedge\eta^6 +\frac{1}{s}\eta^3\wedge\eta^7 -\eta^5\wedge\eta^ 8 \, , \\
\dd\eta^5&= + \, \frac{1}{s}\eta^1\wedge\eta^4 +\frac{1}{s}\eta^2\wedge\eta^7 +\frac{1}{s}\eta^3\wedge\eta^6 +\eta^4\wedge\eta^8 \, , \\
\dd\eta^6&= - \, \frac{1}{s}\eta^1\wedge\eta^7 +\frac{1}{s}\eta^2\wedge\eta^4 -\frac{1}{s}\eta^3\wedge\eta^5 +\eta^7\wedge\eta^8 \, , \\
\dd\eta^7&= + \, \frac{1}{s}\eta^1\wedge\eta^6 -\frac{1}{s}\eta^2\wedge\eta^5 -\frac{1}{s}\eta^3\wedge\eta^4 -\eta^6\wedge\eta^8 \, , \\
\dd\eta^8&= - \, 6 \, \eta^4\wedge\eta^5 + \, 6 \, \eta^6\wedge\eta^7 \, .
\end{split}
\end{align}
It is again easy to check that \eqref{eq:derivativesofetasimproved} and \eqref{eq:derivativesofomegasimproved} are satisfied and the coframe describes the squashed 3-Sasakian structure of the Aloff-Wallach space.

\bigskip

\section{Instanton connections}
\label{sec:InstantonConnections}

The construction of heterotic G$_2$ systems requires the existence of G$_2$-instanton connections on certain vector bundles. In this section we introduce instanton connections on 3-Sasakian manifolds and their squashed deformations, which will play a role in the new solutions we present later.

\subsection{Canonical connection}
\label{sec:CanonicalConnection}

Homogeneous manifolds are equipped with a natural bundle and connection, as explained for example in \cite{Harland:2010ix}. Let $G$ be a Lie group, $H$ a closed Lie subgroup and let $\mathfrak{g}$ and $\mathfrak{h}$ be their Lie algebras. Theorem 11.1 from \cite{kobayashi1963foundations} states that if the coset $G/H$ is reductive then the $\mathfrak{h}$-component of the Maurer-Cartan one-form of $G$ \eqref{eq:maurercartanform} defines a connection in the principal $H$-bundle $G\longrightarrow G/H$ which is left-invariant under the $G$-action. We call this the  \emph{canonical connection on the principal bundle} and write it as
\begin{equation}
\label{eq:bundlecanonicalcon}
A_\text{can}=\sum_a I_a\otimes e^a \, ,
\end{equation}
where $\{e^a\}$ are the vertical one-forms and $\{I_a\}$ is the associated dual basis of $\mathfrak{h}$. Note the canonical connection does not depend on the coset metric so for the case of squashed 3-Sasakian manifolds it is well-defined for all values of the squashing parameter $s$.

For applications to heterotic G$_2$ systems we are interested in considering a representation of the group $H$ and work with the canonical connection \emph{on the associated vector bundle}. The reductiveness of the coset provides a natural representation of $H$ as follows: write $\mathfrak{g}=\mathfrak{h}\oplus\mathfrak{m}$ where $\mathfrak{m}$ is identified with the tangent space at the identity of $G/H$. The elements of $\mathfrak{h}$ act on $\mathfrak{m}$ via the adjoint action, and the condition $[\mathfrak{h},\mathfrak{m}]\subset\mathfrak{m}$ ensures that the action defines a representation of $\mathfrak{h}$ in $\mathfrak{m}$. This provides a representation of $H$ whose associated vector bundle is the tangent space of the coset $G/H$. 

The existence of a canonical connection on the tangent bundle is not exclusive to homogeneous manifolds. As shown in \cite{Harland:2011zs} all 3-Sasakian manifolds have a canonical connection constructed using the forms\footnote{Note our conventions differ slightly from \cite{Harland:2011zs}.} preserved by the Sp(1)-structure of the 3-Sasakian manifold
\begin{equation}
\label{formP}
P=\frac{1}{3} \xi^{123}-\frac{1}{3}\sum_{i=1}^3 \xi^i\wedge\omega^i \, , \qquad
Q=*F_1=\frac{1}{6}\sum_{i=1}^3\omega^i\wedge\omega^i \, .
\end{equation}
The Christoffel symbols of the canonical connection on the tangent bundle are
\begin{equation}
\label{eq:eqsofcanonical}
\Gamma\indices{^i_{\mu\nu}}={}^{LC}\Gamma\indices{^i_{\mu\nu}}+3 \, P_{i\mu\nu} \, , \qquad
\Gamma\indices{^\nu_{\mu i}}=-{}^{LC}\Gamma\indices{^i_{\mu\nu}}-3 \, P_{i\mu\nu} \, , \qquad
\Gamma\indices{^\beta_{\mu\alpha}}={}^{LC}\Gamma\indices{^\beta_{\mu\alpha}} \, ,
\end{equation}
where $\mu,\nu\in\{1 \, ,\dots,7$\}, $i\in\{1,2,3\}$, $\alpha,\beta\in\{4,5,6,7\}$, ${}^{LC}\Gamma$ denotes the Christoffel symbols of the Levi-Civita connection and $P$ is the form defined in \eqref{formP}. The torsion is given by
\begin{equation}
\label{eq:torsionHarland}
    T^i=3 \, P_{i\mu\nu}\,\xi^\mu\wedge \xi^\nu \, , \qquad T^\alpha=\frac{3}{2} P_{\alpha\mu\nu}\,\xi^\mu\wedge \xi^\nu \, .
\end{equation}
This connection is also compatible with all the squashed metrics. In the case of squashed homogeneous 3-Sasakian manifolds, we can regard the structure equations \eqref{eq:struceq} of the coset as Cartan structure equations for the canonical connection
\begin{equation}
\label{cartanstructureeqs}
\dd \xi^\mu=-\omega\indices{^\mu_\nu} \wedge \xi^\nu+ \xi^\mu\lrcorner \,  T \, .
\end{equation}
From this we can read the connection one-form of the canonical connection
\begin{equation}
\label{eq:connectiononeformintermsofstrucconst}
\omega\indices{^\mu_\nu}=f^\mu_{a\nu}\, \xi^a \, ,
\end{equation}
as well as the torsion
\begin{equation}
\label{eq:torsionintermsofstrucconst}
\xi^\mu\lrcorner \,  T=-\frac{1}{2}f^\mu_{\nu\rho}\, \xi^\nu\wedge \xi^\rho \, ,
\end{equation}
where $\mu,\nu,\rho=1 \, ,\dots,7$ and $a=8 \, ,\dots,\rank(\mathfrak{g})$. This agrees with \eqref{eq:torsionHarland},
and we note that with our conventions the canonical connection has totally antisymmetric torsion only when $s=\frac{1}{\sqrt{2}}$.\footnote{This is because for $s=\frac{1}{\sqrt{2}}$ the coset metric we choose is proportional to the Killing form.}

\bigskip

We now turn to the question of whether this connection is a G$_2$-instanton. It is shown in \cite{Harland:2011zs} that the canonical connection in tangent space is an Sp(1)-instanton. This is done by proving that the indices of the curvature of the connection have interchange symmetry and then observing that it has $\mathfrak{sp}(1)$-holonomy.
Since the Sp(1)-structure is given by the forms $\lbrace(\xi^i,\omega^i)_{i=1,2,3}\rbrace$ and the G$_2$-structure is determined by \eqref{eq:threeformforsquash}, the $G$-structures are such that $\mathfrak{sp}(1)\in\mathfrak{g}_2 \, $, see \cite{delaOssa:2021cgd}. As a result, the canonical connection in tangent space is in fact a G$_2$-instanton, and this remains true for all values of the squashing parameter. 

Let us show the G$_2$-instanton condition also holds for the canonical connection on the principal bundle \eqref{eq:bundlecanonicalcon}: consider a squashed 3-Sasakian homogeneous manifold.\footnote{Our approach is similar to \cite{Harland:2010ojo} and we show it case-by-case rather than giving a general derivation. Even though canonical connections seem to always satisfy an instanton condition, to the best of our knowledge there is not a general argument for this in the full general case.} The curvature
\begin{equation}
\label{eq:curvatureofgaugebundle}
F_\text{can}=\dd A_\text{can}+\frac{1}{2}[A_\text{can}\wedge A_\text{can}] \, ,
\end{equation}
of the canonical connection \eqref{eq:bundlecanonicalcon} can be computed using the structure equations \eqref{eq:squashedstruceq3} obtaining
\begin{equation}
\label{eq:curvaturecanonicalgeneral}
F_\text{can}=\sum_a I_a\otimes \left( -\frac{1}{s}f^a_{j\alpha}\, \eta^j\wedge \eta^\alpha-\frac{1}{2s^2}\,f^a_{jk} \,\eta^j\wedge \eta^k-\frac{1}{2}f^a_{\alpha\beta}\, \eta^\alpha\wedge \eta^\beta \right),
\end{equation}
where $j,k\in\lbrace 1,2,3\rbrace$, $\alpha,\beta\in\lbrace 4 \, ,\dots,\rank(\mathfrak{m})\rbrace$, $a\in\lbrace \rank(\mathfrak{m})+1 \, ,\dots,\rank(\mathfrak{g})\rbrace$ and $s$ is the squashing parameter. Comparing \eqref{eq:derivativesofetasimproved} and \eqref{eq:squashedstruceq1} we can rewrite the two-forms $\omega^i$ in terms of the structure constants
\begin{equation}
    \label{eq:omegaintermsofstructureconstants}
    \omega^i=-\frac{1}{2 \, s}f^i_{aj}\,\eta^a\wedge \eta^j-\frac{1}{2}f^i_{a\alpha}\,\eta^a\wedge \eta^\alpha-\frac{1}{2 \, s}f^i_{j\alpha}\, \eta^j\wedge \eta^\alpha-\frac{1}{4}f^i_{\alpha\beta}\, \eta^\alpha\wedge \eta^\beta \, .
\end{equation}
In fact, these expressions are greatly simplified in the 7-dimensional case since most of the structure constants vanish, see \eqref{eq:structureconstantssp2} and \eqref{eq:structureconstantssu3}. We have
\begin{equation}
    \label{eq:curvatureandomega}
    F_\text{can}=\sum_a I_a\otimes \left( -\frac{1}{2}f^a_{\alpha\beta}\, \eta^\alpha\wedge \eta^\beta \right), \qquad \omega^i=-\frac{1}{4}f^i_{\alpha\beta}\, \eta^\alpha\wedge \eta^\beta \, ,
\end{equation}
this can be used together with \eqref{eq:dualF1andF2} and \eqref{eq:fourformsquashed} to express the coassociative form in terms of the structure constants
\begin{equation}
\label{eq:dualF1andF2structureconstants}
\psi_s=\frac{1}{96}\sum_{i=1}^3f^i_{\alpha\beta}\,f^i_{\gamma\rho}\, \eta^{\alpha\beta\gamma\rho}-\frac{1}{8}\epsilon_{ijk}\,f^k_{\alpha\beta}\, \eta^{ij\alpha\beta} \, ,
\end{equation}
where $i,j,k\in\lbrace 1,2,3\rbrace$, $\alpha,\beta,\gamma,\rho\in\lbrace 4,\dots,7\rbrace$. The instanton condition $F_\text{can}\wedge\psi=0$ for the canonical connection can then be written in terms of the structure constants as
\begin{equation}
    \label{eq:instantonconditionstructureconstants}
    f^a_{[\alpha\beta}\,f^k_{\gamma\rho]}=0 \, ,
\end{equation}
for all $k\in\lbrace 1,2,3\rbrace$, $\alpha,\beta,\gamma,\rho\in\lbrace 4 \, ,\dots,7\rbrace$, $a\in\lbrace 8 \, ,\dots,\rank(\mathfrak{g})\rbrace$. It is immediate to check this is satisfied for both the squashed 7-sphere and the squashed Aloff-Wallach space, so the canonical connection is a G$_2$-instanton for both of them, for all values of the squashing parameter $s$.

\subsection{Clarke-Oliveira connection}
\label{sec:ClarkeOliveiraConnection}

An additional G${}_2$-instanton connection on squashed 3-Sasakian manifolds is described in \cite{2019arXiv190305526C}. We briefly review the construction and adapt it to our purposes. This involves a straightforward generalization to all values of the squashing parameter $s$.

Any 3-Sasakian manifold $Y$ is described as the total space of an SU(2)-bundle over a Riemannian orbifold $Z$. This bundle can be identified with a bundle of self-dual antisymmetric two-forms on $Z$, and as a result, the Levi-Civita connection of $Z$ induces a connection on the bundle $Y\longrightarrow Z$,
\begin{equation}
\label{eq:connectiononZbeforelift}
a_Z=\sum_{i=1}^3 I_i\otimes \xi^i \, ,
\end{equation}
where $I_i\in\mathfrak{su}(2)$ with $[I_i,I_j]=2 \, \epsilon\indices{_{ij}^k} \, I_k$ for $i,j,k\in\lbrace 1,2,3\rbrace$, and $\xi^i$ denotes the pullback of the 3-Sasakian one-forms to $Z$.\footnote{When the 4-dimensional orbifold $Z$ can be described as a coset, this is just the canonical connection on the $Y\longrightarrow Z$ bundle as described in \cref{sec:CanonicalConnection}.} Consider now the trivial SU(2) bundle over $Y$, and pull back the connection \eqref{eq:connectiononZbeforelift} to the bundle $Y\times\text{SU}(2)\longrightarrow Y$ to obtain\footnote{The main focus of \cite{2019arXiv190305526C} is the study of Spin(7)-instantons on cones over squashed 3-Sasakian manifolds, which makes convenient to choose the vector bundle associated to the fundamental SU(2) representation and to introduce a radial dependence in the connection. We do not make a choice of representation so there is no radial factor present in our version of the construction.}
\begin{equation}
a(x_1,x_2,x_3)=\sum_{i=1}^{3} x_i \,  I_i\otimes\xi^i \, .
\end{equation}
Here $x_i\in\R$ are free parameters to be fixed later by the instanton condition. As we squash the metric, the connection in terms of the orthonormal coframe \eqref{eq:coframe} is given by
\begin{equation}
\label{eq:clarkeoliveiraconnection}
a(x_1,x_2,x_3)=\sum_{i=1}^{3} \frac{x_i}{s} I_i\otimes\eta^i \, .
\end{equation}
The curvature of the connection
\begin{equation}
\label{eq:curvatureofgaugebundle2}
F=\dd a+\frac{1}{2}[a\wedge a]
\end{equation}
changes with the squashing and can be computed using \eqref{eq:derivativesofetasimproved}, obtaining
\begin{equation}
\label{eq:ClarkeOliveiracurvaturebeforeinstanton}
F(x_1,x_2,x_3)=\sum_{i=1}^{3} \left(2 \, x_i \, \omega^i+\sum_{j,k=1}^{3}\epsilon\indices{^i_j_k}\,\frac{1}{s^2}\,\left(-x_i+x_jx_k\right)\eta^j\wedge \eta^k\right)\otimes I_i \, .
\end{equation}
The G${}_2$-instanton condition $F\wedge\psi=0$ for this connection reduces to three equations
\begin{equation}
\left(2-\frac{1}{s^2}\right)x_i+\frac{1}{s^2}x_j \, x_k=0, \qquad \text{where } \ i,j,k\in\lbrace 1,2,3\rbrace \ \text{ and } \ i\neq j\neq k \, .
\end{equation}
Imposing these equations leads to two different non-zero kind of instantons. For a fixed squashing of $s=\frac{1}{\sqrt{2}}$ setting any two of the $x_i$ parameters to zero gives a solution for any value of the remaining $x_i \, $. Therefore, there are three 1-parameter families of G$_2$-instantons for $s=\frac{1}{\sqrt{2}}$. Unfortunately, the curvature of these connections is such that they do not provide solutions of the heterotic G$_2$ system.

The other group of instantons is defined for any value of $s$ and can be found by setting the $x_i$ parameters to $\pm(1-2s^2)$ with the following restriction: either we choose all signs positive\footnote{For the nearly-parallel case $s=\frac{1}{\sqrt{5}}$ this connection has $x_i=\frac{3}{5}$ for $i=1,2,3$ and it is precisely the instanton connection described in Example 2 of \cite{2019arXiv190305526C}.}, or we choose one of them positive and the rest negative. This provides four different G$_2$-instantons for all values of $s$ which we will use to construct new solutions of the heterotic system. Note that all of them reduce to the trivial flat connection for $s=\frac{1}{\sqrt{2}}$.

\subsection{Tangent bundle instantons}
\label{sec:tangentbundleinstantons}

In this section we focus on the tangent bundle and describe a one-parameter family of G$_2$-instantons. We briefly review how to compute the connection one-form and fix our notation, which agrees with \cite{Nakahara:2003nw}. Given a connection on the tangent bundle, consider the Cartan structure equations in the orthonormal coframe
\begin{equation}
\label{eq:cartanfororthonormalbasis}
\dd \eta^\mu=-\omega\indices{^\mu_\nu} \wedge \eta^\nu+ \eta^\mu\lrcorner T \, ,
\end{equation}
where $\omega\indices{^\mu_\nu}=\Gamma\indices{^\mu_{\alpha\nu}} \, \eta^\alpha$ is the connection one-form and $T$ is the torsion tensor, with $\mu,\nu,\rho=1 \, ,\dots,7$ and $\alpha=1 \, ,\dots,\rank(\mathfrak{g})$.
The torsion is antisymmetric in the last two indices, $T\indices{^\mu_{\nu\rho}}=-T\indices{^\mu_{\rho\nu}}$, and $\eta^\mu\lrcorner  \, T=T^\mu=\frac{1}{2}T\indices{^\mu_{\nu\rho}}\eta^\nu\wedge\eta^\rho$.

Manipulating the Cartan equations \eqref{eq:cartanfororthonormalbasis} we can write the connection one-form as
\begin{equation}
\label{eq:oneformfinalformula}
\omega\indices{^\mu_{\nu}}={}^{LC}\omega\indices{^\mu_{\nu}}+K\indices{^\mu_{\rho\nu}}\eta^\rho \, .
\end{equation}
where ${}^{LC}\omega$ is the connection one-form of the Levi-Civita connection, given by
\begin{equation}
\label{eq:levicivitaformula}
{}^{LC}\Gamma_{\mu\nu\rho}=-\frac{1}{2}\left[(\dd \eta_\mu)_{\nu\rho}-(\dd \eta_\nu)_{\rho\mu}+(\dd \eta_\rho)_{\mu\nu}\right] \, ,
\end{equation}
and $K$ is the \emph{contorsion tensor}
\begin{equation}
\label{eq:contorsioneq}
K_{\mu\nu\rho}=\frac{1}{2}\left(T_{\mu\nu\rho}+T_{\nu\mu\rho}+T_{\rho\mu\nu}\right) \, ,
\end{equation}
which is antisymmetric in the first and last indices, ensuring the connection we define is compatible with the metric. Indices are raised and lowered with the orthonormal metric. Note that since the Levi-Civita connection is determined by the structure equations, \eqref{eq:oneformfinalformula} shows that connections are completely specified by the choice of torsion tensor.

Every manifold with a G$_2$-structure \cite{2003math......5124B} has a 2-parameter family of metric connections preserving the G$_2$-structure. The torsion can in fact be written explicitly and it is given by \cite{delaOssa:2017pqy}
\begin{align*}
\frac{1}{2}T_{\mu\nu\rho}(a,\beta)=\frac{1}{12}\tau_0 \, \varphi_{\mu\nu\rho}-\frac{1}{6}\tau_{2\sigma[\nu} \, \varphi\indices{_{\rho]\mu}^\sigma}+a \, \tau_{3\mu\nu\rho}+\frac{1}{4}(1+2 \, a)S\indices{_\mu^\sigma} \, \varphi_{\nu\rho\sigma}+\\
+\frac{1}{3}(\beta-1)(\tau_1\lrcorner \, \psi)_{\mu\nu\rho}+\frac{2}{3}(1+2 \, \beta)\tau_{1[\nu} \, g_{\rho]\mu} \, ,
\end{align*}
where $S_{\mu\nu}=\frac{1}{4}\varphi\indices{^{\rho\sigma}_{(\mu}} \, (\tau_3){}_{\nu)\rho\sigma}$ and $a,\beta\in\R \, $. For squashed 3-Sasakian manifolds the torsion classes $\tau_2$ and $\tau_1$ vanish, see \eqref{eq:tau1} and \eqref{eq:tau2}, so the family reduces to one parameter\footnote{The choice of parameter $a=-\frac{1}{2}$ corresponds the unique G$_2$-compatible connection with totally antisymmetric torsion \eqref{eq:uniqueantitorsion}.}
\begin{equation}
\label{eq:torsiongeneralg2}
\frac{1}{2}T_{\mu\nu\rho}(a)=\frac{1}{12}\tau_0 \, \varphi_{\mu\nu\rho}+a \, \tau_{3\mu\nu\rho}+\frac{1}{4}(1+2 \, a)S\indices{_\rho^\sigma}\varphi_{\sigma\mu\nu}.
\end{equation}
Using the torsion classes computed in \eqref{eq:tau0} and \eqref{eq:tau3} we obtain the torsion of the family\footnote{For the nearly-parallel case $s=\frac{1}{\sqrt{5}}$ the vanishing of $\tau_3$ reduces the family to a single connection which is the one with totally antisymmetric torsion.} of G${}_2$-compatible connections for squashed 3-Sasakian manifolds in terms of $s$ and $a$
\begin{align}
T_{ijk} &=\left[(2+20 \, a)s-4 \, a\frac{1}{s}\right]\varphi_{ijk}, \\
T_{iab} &=2 \, s \, \varphi_{iab}, \\
T_{aib} &=\left[-\left(\frac{1}{2}+5 \, a\right)s+\left(\frac{1}{2}+a\right)\frac{1}{s}\right]\varphi_{aib},
\end{align}
where $i,j,k=1,2,3$ and $a,b=4,5,6,7$. The curvature of this family of connections has to be computed in a case-by-case basis since it depends on the Levi-Civita connection of the manifold, recall \eqref{eq:oneformfinalformula}. We have done this for the 7-dimensional squashed homogeneous 3-Sasakian manifolds in \cref{sec:ConnectionCurvatureg2compatible} and used it to check the G${}_2$-instanton equations $F\wedge\psi=0$.

We find that the connections with torsion \eqref{eq:torsiongeneralg2} are G${}_2$-instantons for all values of the parameters $s$ and $a$ for both the squashed 7-sphere and the squashed Aloff-Wallach space.\footnote{\label{foot:canisinfamily} This family includes the canonical connection with the representation on the tangent bundle we described in \cref{sec:CanonicalConnection}, which is recovered for the choice of parameter $a(s)=\frac{1+s^2}{2-10s^2}$. On the other hand, the Clarke-Oliveira connection does not belong to this 1-parameter family.}

We finish with a comment on other possible instantons: note the heterotic G${}_2$-system introduced in \cref{sec:Heteroticg2Systems} does not require the tangent space instanton to be compatible with the G$_2$-structure. Therefore, we could look for instantons outside the family \eqref{eq:torsiongeneralg2}. We have explored this possibility for the homogeneous cases and several families of instantons can be found. Nevertheless, the curvatures of these connections are quite involved and imposing the heterotic Bianchi identity \eqref{eq:heteroticBianchiidentity} becomes difficult. Therefore, we will not use these instantons in our solutions. It would be interesting to verify if it is indeed possible to use these other connections to obtain solutions.

\bigskip

\section{The heterotic Bianchi identity}\label{sec:HeteroticBI}

As described in \cref{sec:Heteroticg2Systems}, a heterotic G$_2$ system includes an instanton connection $A$ on a vector bundle $V$ as well as an instanton connection $\Theta$ on the tangent bundle. The curvatures of these connections must satisfy the heterotic Bianchi identity
\begin{equation}
\dd H=\frac{\alpha'}{4}\big(\tr (F_A\wedge F_A) -\tr (R_\Theta\wedge R_\Theta)\big) \, ,
\qquad H = T \, .
\end{equation}
In \cref{sec:InstantonConnections} we have presented several instanton connections on different bundles over 7-dimensional squashed homogeneous  3-Sasakian manifolds. In order to use these instantons to construct heterotic G$_2$ systems and verify they satisfy the heterotic Bianchi identity, in this section we compute the terms of the form $\tr (F\wedge F)$.

Considering the curvature as a Lie algebra-valued two-form, the trace is taken over a product of Lie algebra generators $\tr (I_a I_b)$. The value of this trace will depend on the Lie algebra representation associated to the vector bundle where the connection is defined. Some of the connections introduced in \cref{sec:InstantonConnections} are defined on principal bundles, so a representation of the gauge group has to be chosen to obtain an associated vector bundle. We will explicitly see how the value of $\tr (F\wedge F)$ depends on this choice.

\subsection{Canonical connection}
\label{sec:tracecanonicalconnection}

Recall the canonical connection of a homogeneous manifold was described in \cref{sec:CanonicalConnection}
\begin{equation}
A_\text{can}=\sum_a I_a\otimes e^a \, .
\end{equation}
Using the structure equations \eqref{eq:generalstructureequation}, we computed the curvature of the canonical connection for 7-dimensional squashed homogeneous 3-Sasakian manifolds
\begin{equation}
    F_\text{can}=\sum_a I_a\otimes \left( -\frac{1}{2}f^a_{\alpha\beta}\, \eta^\alpha\wedge \eta^\beta \right) \, ,
\end{equation}
and now it is immediate to compute
\begin{equation}
\label{eq:tracegeneralcanonical}
    \tr( F_\text{can}\wedge F_\text{can})=\sum_{a,b} \tr(I_aI_b) \otimes \left( \frac{1}{4}f^a_{\alpha\beta}\,f^b_{\gamma\rho} \ \eta^{\alpha\beta\gamma\rho} \right) \, ,
\end{equation}
where $\alpha,\beta,\gamma,\rho\in\lbrace 4,\dots,7\rbrace$, $a\in\lbrace 8 \, ,\dots,\rank(\mathfrak{g})\rbrace$. As we have said before, a representation of the gauge group has be chosen to compute explicitly $\tr(I_aI_b)$, so we need 
to distinguish between the 7-sphere and the Aloff-Wallach space.

\subsubsection{The squashed 7-sphere}

\label{sec:canonicalconnforthe7sphere}

The canonical connection on the squashed 7-sphere is defined on a principal SU(2)-bundle. Representations of SU(2) are in bijective correspondence with representations of $\mathfrak{su}(2)$ and are well understood, see for example \cite{Fulton1991RepresentationTA} or \cite{Yamatsu:2015npn}. Complex representations of $\mathfrak{su}(2)$ are classified by a single Dynkin label $(m)$, so that $(m)$ corresponds to the $(m+1)$-dimensional representation.

The $\mathfrak{su}(2)$ generators $\lbrace I_8 \, , I_9 \, , I_{10} \rbrace$ satisfy the commutation relations $[I_a,I_b]=2 \, \epsilon\indices{_{ab}^c} \, I_c$. The trace of a product of generators of a representation is directly related to the \emph{Dynkin index} of the representation. With our conventions, the index of the representation $(m)$ is given by
\begin{equation}
    c(m)=\frac{1}{3}m(m+1)(m+2)
\end{equation}
and the trace of the  representation $(m)$ is
\begin{equation}
    \tr(I_a I_b)=-c(m) \, \delta_{ab} \, .
\end{equation}
Consider now a general finite-dimensional representation of SU(2) and the canonical connection in the associated vector bundle $V$. The corresponding $\mathfrak{su}(2)$ representation is a direct sum of $k$ irreducible representations with Dynkin labels $m_1,\dots,m_k$. Since the Dynkin index is additive and using the structure constants \eqref{eq:structureconstantssp2}, we obtain from \eqref{eq:tracegeneralcanonical}
\begin{equation}
\label{eq:tracecanonicalsphere}
    \tr( F_\text{can}\wedge F_\text{can})=24\left( c(m_1)+\cdots+c(m_k) \right)*_sF_1 \, ,
\end{equation}
where $*_sF_1=\eta^{4567}$ was defined in \eqref{eq:dualF1andF2}.

Some representations deserve a special mention. First of all, note that the smallest value of the trace is obtained for the fundamental representation, where
\begin{equation}
    \tr( F_\text{can}\wedge F_\text{can})=48*_sF_1 \, .
\end{equation}
whereas the adjoint representation gives
\begin{equation}
    \tr( F_\text{can}\wedge F_\text{can})=192*_sF_1 \, .
\end{equation}
We stressed in \cref{sec:CanonicalConnection} that the canonical connection has a natural representation on the tangent bundle due to the coset $G/H$ being reductive. The representation is described by the adjoint action of $H$ on $G/H$, so the matrix representation of the generators is given by the structure constants
\begin{equation}
    \left( I_a \right)_{\mu\nu}=f^\mu_{a\nu} \, ,
\end{equation}
the explicit matrices can be found in \cref{sec:ExpRepMatSp1AdjAct} and they satisfy
\begin{equation}
    \tr(I_a I_b)=-4\delta_{ab} \, .
\end{equation}
Therefore, the canonical connection in the tangent bundle representation has
\begin{equation}
\label{eq:tracecanonicaltangentsphere}
    \tr(F_\text{can}\wedge F_\text{can})=96 *_sF_1 \, .
\end{equation}
Equivalently, the tangent bundle representation of the canonical connection can be described using Cartan structure equations as we showed in \eqref{eq:connectiononeformintermsofstrucconst}, obtaining the Christoffel symbols \eqref{eq:eqsofcanonical} of \cite{Harland:2011zs}.

\subsubsection{The squashed Aloff-Wallach space}

In the case of the squashed Aloff-Wallach space, the canonical connection is on a principal U(1) bundle.
Since U(1) is abelian, all irreducible complex representations are 1-dimensional and described by maps from the circle to itself classified by an integer $q$
\begin{equation}
    \text{U}(1)\ni e^{i\theta} \longmapsto e^{iq\theta}\in\text{U}(1) \, .
\end{equation}
The Lie algebra generator is then given by
\begin{equation}
    I_8=i \, q \, ,
\end{equation}
which results in
\begin{equation}
    \tr(I_8 I_8)=-q^2 \, .
\end{equation}
On the other hand, we can consider irreducible (non-trivial) real representations. These are all 2-dimensional, taking values in SO(2) and classified by a positive integer $p$
\begin{equation}
    \text{U}(1)\ni e^{i\theta} \longmapsto \begin{pmatrix}
    \cos{(p\theta)} & -\sin{(p\theta)} \\
    \sin{(p\theta)} & ~\cos{(p\theta)}
    \end{pmatrix} \in\text{SO}(2) \, .
\end{equation}
The Lie algebra generator is then given by
\begin{equation}
    I_8=\begin{pmatrix}
0 &-p \\
p & ~0 
\end{pmatrix},
\end{equation}
which results in
\begin{equation}
    \tr(I_8 I_8)=-2 \, p^2 \, .
\end{equation}
Let us consider now a general finite-dimensional representation of U(1) and the associated vector bundle $V$ together with the corresponding canonical connection. The representation is given by a direct sum of irreducible representations. Suppose we have $k$ irreducible complex representations with indices $p_1,\dots,p_k$ and $\ell$ irreducible real representations with indices $q_1,\dots,q_\ell$. Using the structure constants \eqref{eq:structureconstantssu3}, we obtain from \eqref{eq:tracegeneralcanonical}
\begin{equation}
\label{eq:tracecanonicalaloffwallach}
    \tr( F_\text{can}\wedge F_\text{can})=72\left( q_1^2+\cdots+q_k^2 \right)*_sF_1+144\left( p_1^2+\cdots+p_\ell^2 \right)*_sF_1 \, ,
\end{equation}
where $*_sF_1=\eta^{4567}$ was defined in \eqref{eq:dualF1andF2}.

In this case, the smallest value possible for the trace term is obtained for a single complex representation
\begin{equation}
\label{eq:traceAloffcanonicalgeneral}
    \tr( F_\text{can}\wedge F_\text{can})=72\,*_s F_1 \, .
\end{equation}
Interestingly, a real representation with index $q$ gives the same trace value as a direct sum of two complex representations with indices $\pm q$. This means different representations can be used in exactly the same way to obtain solutions. 

Consider now the natural representation of the canonical connection on tangent space described in \cref{sec:CanonicalConnection}. The matrix representation of the generator is given by
\begin{equation}
    \left( I_8 \right)_{\mu\nu}=f^\mu_{8\nu} \, ,
\end{equation}
the matrix is explicitly written in \cref{sec:ExpRepMatSp1AdjAct} and satisfies
\begin{equation}
    \tr(I_8 I_8)=-4 \, .
\end{equation}
The canonical connection in the tangent bundle representation gives
\begin{equation}
\label{eq:tracecanonicaltangentAloffWallach}
    \tr(F_\text{can}\wedge F_\text{can})=288 *_sF_1 \, .
\end{equation}
It is clear from the explicit matrix form that the tangent bundle representation is a direct sum of the trivial representation on the coordinates $1,2,3$ and two real representations with index $p=1$ on the coordinates $4,5$ and $7,6$. This can be used to recover the trace term from the general formula \eqref{eq:traceAloffcanonicalgeneral}. Alternatively, one can describe the canonical connection using Cartan structure equations as in \eqref{eq:connectiononeformintermsofstrucconst}, and the computation of the curvature becomes very simple since U(1) is abelian.

\subsection{Clarke-Oliveira connection}
\label{sec:traceclarkeoliveira}

As found in \cref{sec:ClarkeOliveiraConnection} the Clarke-Oliveira connection
\begin{equation}
a(x_1,x_2,x_3)=\sum_{i=1}^{3} \frac{x_i}{s} I_i\otimes\eta^i \, ,
\end{equation}
is an instanton for all values of the squashing parameter $s$ if we set the $x_i$ parameters to $\pm(1-2s^2)$ with the following restriction: either we choose all signs positive, or we choose one of them positive and the rest negative.

Since the Clarke-Oliveira connection is defined on an SU(2)-principal bundle, we can choose representations in the same way as for the canonical connection on the squashed 7-sphere in \cref{sec:canonicalconnforthe7sphere}: the $(m+1)$-dimensional complex representations of $\mathfrak{su}(2)$ is denoted by the Dynkin label $(m)$, and the trace of its generators is given by
\begin{equation}
    \tr(I_i I_j)=-c(m)\delta_{ij} \, ,
\end{equation}
where $c(m)$ is the Dynkin index of the representation, given in our conventions by
\begin{equation}
    c(m)=\frac{1}{3}m(m+1)(m+2) \, .
\end{equation}
For every representation the nonzero contribution to $\tr(I_i I_j)$ comes from the terms with $i=j$, so from the curvature \eqref{eq:ClarkeOliveiracurvaturebeforeinstanton} we obtain, for all four choices of $x_i$ we have previously indicated
\begin{equation}
\label{eq:clarkeoliveiratracewedgeunfinished}
    \tr(F\wedge F)=\sum_{i=1}^3 8(1-2 \, s^2)^2\left( *_s F_1-\epsilon\indices{^i_j_k}\,\omega^i\wedge\eta^j\wedge\eta^k \right)\tr(I_i I_i) \, ,
\end{equation}
Take a finite-dimensional representation of SU(2), the corresponding $\mathfrak{su}(2)$ representation is a direct sum of $k$ irreducible representations with Dynkin labels $m_1,\dots,m_k$. Substituting $\tr(I_i I_i)$ for this representation in \eqref{eq:clarkeoliveiratracewedgeunfinished} we obtain the general formula
\begin{equation}
\label{eq:traceclarkeoliveira}
    \tr( F\wedge F)=-8(1-2 \, s^2)^2\left( c(m_1)+\cdots+c(m_k) \right)\left( 3*_s F_1-2*_s F_2 \right) \, ,
\end{equation}
where $*_s F_1$ and $*_s F_2$ were defined in \eqref{eq:dualF1andF2}.

The Clarke-Oliveira connection is obtained via pullback by regarding the squashed 3-Sasakian manifold as the total space of an SU(2)-bundle. Therefore, the $\mathfrak{su}(2)$ algebra has a natural adjoint action on the tangent bundle of the 3-Sasakian manifold which can be used to construct a representation of the Clarke-Oliveira connection. For homogeneous manifolds this is analogous to the tangent bundle representation of the canonical connection and the explicit matrices are given by the structure constants
\begin{equation}
    \left( I_i \right)_{\mu\nu}=f^\mu_{i\nu} \, .
\end{equation}
The matrices are the same for both the squashed 7-sphere and the squashed Aloff-Wallach space and they can be found in \cref{sec:ExpRepMatCOconn}. They satisfy
\begin{equation}
\tr(I_i I_j)=-12 \, \delta_{ij} \, ,
\end{equation}
so that the Clarke-Oliveira connection in the tangent bundle satisfies
\begin{equation}
\label{eq:traceclarkeoliveiratangent}
    \tr( F\wedge F)=-96(1-2 \, s^2)^2\left( 3*_s F_1-2*_s F_2 \right).
\end{equation}

\subsection{Tangent bundle instantons}
\label{sec:tracetangentbundle}

In \cref{sec:tangentbundleinstantons} we introduced a 1-parameter family of instantons on the tangent bundle of the squashed 7-sphere and the squashed Aloff-Wallach space consisting on the most general metric connections compatible with the G$_2$-structures. The explicit expressions of the connections and curvatures can be found in \cref{sec:ConnectionCurvatureg2compatible} and can be used to compute the term $\tr(F\wedge F)$ for both manifolds. For simplicity, let us denote
\begin{equation}
    \kappa(a,s)=(1+10 \, a)s+(1-2 \, a)\frac{1}{s} \, ,
\end{equation}
for the squashed 7-sphere we then find
\begin{equation}
\label{eq:traceofgeneralg2connection}
    \tr(F\wedge F)= -72 \, s^2 \left( \kappa(a,s)^2 -\frac{4}{3s^2} \right)*_s F_1 -12 \, s \ \kappa(a,s)^2\left(\kappa(a,s)-\frac{2}{s}\right)*_s F_2 \, .
\end{equation}
whereas for the squashed Aloff-Wallach space we obtain
\begin{equation}
\label{eq:traceofgeneralg2connectionAloffWallach}
    \tr(F\wedge F)= -72 \, s^2 \left( \kappa(a,s)^2 -\frac{4}{s^2} \right)*_s F_1 -12 \, s \ \kappa(a,s)^2\left(\kappa(a,s)-\frac{2}{s}\right)*_s F_2 \, .
\end{equation}

\bigskip

\section{New solutions}
\label{sec:NewSolutions}

In this section we provide new solutions to the heterotic G$_2$ system introduced in \cref{sec:Heteroticg2Systems}. This means we have to specify a quadruple
\begin{equation}
    [(Y,\varphi),(V,A),(TY,\Theta),H] \, .
\end{equation}
The 7-dimensional manifold with an integrable G$_2$-structure $(Y,\varphi)$ is a squashed homogeneous 3-Sasakian manifold with squashed metric \eqref{eq:squashedmetric} and G$_2$-structure given by \eqref{eq:threeformforsquash}, for a certain value of the squashing parameter $s$ that we do not fix for now. The flux $H$ is then completely determined and equal to the unique totally antisymmetric torsion given by \eqref{eq:uniqueantitorsion}. Using \eqref{eq:exterdersF1andF2} and \eqref{eq:extderG2squashed} we can compute
\begin{equation}
\label{eq:exteriorderivativeflux}
    \dd H=24 \, s^2*_s F_1+8(1-2 \, s^2)*_s F_2 \, .
\end{equation}

For an instanton connection $\Theta$ on the tangent bundle $TY$, we consider either the one-parameter family we introduced in \cref{sec:tangentbundleinstantons}, or the tangent bundle representations of either the canonical connection or the Clarke-Oliveira connection.

For the vector bundle $V$ and the instanton connection $A$, one option is to choose the canonical connection or the Clarke-Oliveira connection with a representation of the gauge bundle where they were defined. Another option is to take the gauge vector bundle to be the tangent bundle and use the one-parameter family of instantons. We would like to emphasize from the beginning that this is not the so called standard embedding \cite{Candelas:1985en} because we will choose different connections on each bundle.

The final step to solve the heterotic G$_2$ system is to impose the heterotic Bianchi identity \eqref{eq:heteroticBianchiidentity}
\begin{equation}
    \dd H=\frac{\alpha'}{4}\left(\tr(F\wedge F)-\tr(R_\Theta\wedge R_\Theta)\right) \, ,
\end{equation}
with positive string parameter $\alpha'>0$.

The elements $\dd H$, $\tr(F\wedge F)$ and $\tr(R_\Theta\wedge R_\Theta)$ consist of a sum of two terms proportional to $*_s F_1$ and $*_s F_2$, as defined in \eqref{eq:dualF1andF2}. By grouping the coefficients of $*_s F_1$ and $*_s F_2$ in the heterotic Bianchi identity we obtain two independent equations for $s$, $\alpha'$ and any additional coefficients of the connections. These equations impose non-trivial relations between the G$_2$-structure and the curvature of the instantons. In fact, for some choices of connections it is not possible to find a solution, whereas for the rest, the valid ranges of $s$ and $\alpha'$ are restricted. We summarize our findings in \cref{tab:solutions}, which apply to both the squashed 7-sphere and the squashed Aloff-Wallach space.

\begin{table}[h]
{
\centering
\begin{tabular}{|C{0.22\textwidth}||C{0.22\textwidth}|C{0.22\textwidth}|C{0.22\textwidth}|}
\hline
\diagbox[innerwidth=0.22\textwidth,height=2.9\line]{$(V,A)$}{$(TY,\Theta)$} &
Canonical \newline connection &
Clarke-Oliveira connection &
One-parameter family \\
\hline
\hline
Canonical \newline connection &
Solutions only for $s=\frac{1}{\sqrt{2}}$, fixed $\alpha'$ &
Solutions for isolated values of $s$ and $\alpha'$. &
Solutions in different ranges of $s$ with $\alpha'$ determined. \\
\hline
Clarke-Oliveira connection &
No solution &
No solution &
Solutions in different ranges of $s$ with $\alpha'$ determined. \\
\hline
One-parameter family &
No solution &
Solutions in different ranges of $s$ with $\alpha'$ determined. &
Solutions with arbitrary $s$ and $\alpha'$ within a certain range. \\
\hline
\end{tabular}
\caption{Summary of solutions obtained for squashed homogeneous 7-dimensional 3-Sasakian manifolds in terms of the choice of instanton connections.
}
\label{tab:solutions}
}
\end{table}

We give details of these solutions for both the squashed 7-sphere and the squashed Aloff-Wallach space in the following sections, following the order of \cref{tab:solutions} row by row.

Let us first comment on some general features of our solutions. First of all, we are not able to find solutions for the nearly-parallel squashed metric for which $s=\frac{1}{\sqrt{5}}$. One of the reasons is that the one-parameter family of connections in the tangent bundle collapses to a single connection which is in fact the canonical connection in tangent space. Therefore, the set of instanton connections at our disposal for this particular value of $s$ is much smaller.

On the other hand, we do find solutions for all other values of $s>0$. It is interesting to analyse the behaviour of the solutions close to the limits $s=0$ and $s=\infty$. Since the nonzero elements of the Ricci tensor \cite{2010JGP....60..326A} are given by
\begin{equation}
    \mathcal{R}_{ii}=6(2-s^2), \qquad \mathcal{R}_{\alpha\alpha}=\frac{2+4 \, s^4}{s^2} \, ,
\end{equation}
where $i=1,2,3$ and $\alpha=4,5,6,7$, we see the Ricci tensor blows up both for $s=0$ and $s=\infty$. Therefore, these are singular limits and we do not have well-defined solutions for them. Nevertheless, we have a geometrical understanding of the origin of the singularities.

When $s\rightarrow 0$ the $\text{SU}(2)\simeq\Sc^3$ fibres of the squashed 3-Sasakian manifold are shrunk to zero size, obtaining a singular space. In this limit $\dd H\rightarrow 0$ and we always find $\alpha'\rightarrow 0$. This means that the string parameter vanishes as the fibres shrink. We stress that we obtain well-defined solutions only while $s>0$.

The case $s\rightarrow\infty$ corresponds to a large volume limit. The radius of the $\text{SU}(2)\simeq\Sc^3$ fibres of the squashed 3-Sasakian manifold increases and in the limit the manifold decompactifies. Since the G$_2$-structure becomes singular our solutions are only well-defined while $s<\infty$. In this limit $\dd H$ blows up and typically so do the curvatures of the connections involved in the heterotic Bianchi identity \eqref{eq:heteroticBianchiidentity}. Hence, the asymptotic behaviour of $\alpha'$ depends on the particular solution and we find that $\alpha'$ can tend to $0$, $\infty$ or a fixed constant. See more details about this behaviour in \cref{Summary}.

As mentioned earlier, another aspect all the solutions we have constructed have in common is that the $\tau_0$ torsion class \eqref{eq:tau0} is always nonzero. Therefore, the 3-dimensional spacetime emerging from these compactifications is Anti-de Sitter space. Note that the AdS$_3$ curvature is proportional to $\tau_0^2$ and therefore tends to $\infty$ both in the limits $s\rightarrow 0$ and $s\rightarrow\infty$. The minimum value of the curvature is achieved for $s=\frac{1}{\sqrt{2}}$.

Finally, we explain how the solutions can be described when the one-parameter family of instantons of \cref{sec:tangentbundleinstantons} is chosen. These solutions are richer, as was to be expected since we are introducing an extra parameter $a$ in our equations. As we see from \eqref{eq:traceofgeneralg2connection} and \eqref{eq:traceofgeneralg2connectionAloffWallach}, the contribution of this one-parameter family of connections to the heterotic Bianchi identity is cubic in $a$. Thus, the heterotic Bianchi identity can be rewritten as a cubic equation for $a$ with coefficients depending on $s$. We list the coefficients of the cubic equations in \cref{sec:cubiceqs}.

This can be solved explicitly: for each value of $s$ we have up to three different real solutions $a(s)$ of the cubic equation, depending on the sign of the discriminant. For later convenience we define the following functions, motivated by the discriminants in the squashed 7-sphere and the squashed Aloff-Wallach case respectively
\begin{equation}
\label{eq:fandg}
    f(s)=-\frac{4}{9} \, \frac{32 \, s^6-96 \, s^4+78s \, ^2-23}{2 \, s^2-1} \, , \qquad g(s)=-\frac{4}{27} \, \frac{32 \, s^6-96 \, s^4+42 \, s^2-5}{2 \, s^2-1} \, .
\end{equation}
From $a(s)$ we can then obtain the value of $\alpha'$ that solves the heterotic Bianchi identity and ensure it is positive. In most cases the precise expressions of the solutions are not particularly illuminating and we will not show them explicitly.

From the point of view of physics, we are interested in solutions for which the string parameter $\alpha'$ is small. We will highlight these solutions along our presentation.

\subsection{Canonical connection on the vector bundle}

If we want to use the canonical connection on the vector bundle we need to make a choice of representation for the principal bundle. This differs slightly depending on the homogeneous space we choose as we now recall, see \cref{sec:tracecanonicalconnection} for further details.

The canonical connection for the squashed 7-sphere is defined on an SU(2)-bundle. We choose an arbitrary $\mathfrak{su}(2)$ representation, which is given as a direct sum of $k$ irreducible representations with Dynkin labels $m_1,\dots,m_k$. Let us denote for simplicity
\begin{equation}
    c=\left( c(m_1)+\cdots+c(m_k) \right),
\end{equation}
note $c$ can take any natural even value. Then the contribution of the canonical connection to the heterotic Bianchi identity \eqref{eq:tracecanonicalsphere} is given by
\begin{equation}
    \tr( F_\text{can}\wedge F_\text{can})=24 \, c*_s F_1 \, .
\end{equation}
The canonical connection on the squashed Aloff-Wallach space is defined on a U(1) bundle. An arbitrary representation is given as a direct sum of $k$ irreducible complex representations with indices $p_1,\dots,p_k$ and $\ell$ irreducible real representations with indices $q_1,\dots,q_\ell$. Let us denote for simplicity
\begin{equation}
    q=\left( q_1^2+\cdots+q_k^2 \right)+2\left( p_1^2+\cdots+p_\ell^2 \right),
\end{equation}
note $q$ can take any natural value. Then the contribution of the canonical connection to the heterotic Bianchi identity \eqref{eq:tracecanonicalaloffwallach} is given by
\begin{equation}
    \tr( F_\text{can}\wedge F_\text{can})=72 \, q*_sF_1 \, .
\end{equation}
We study the available solutions depending on the choice of tangent bundle instanton.

\subsubsection{Canonical connection on the tangent bundle}
\label{subsec:cancan}

We only find a single solution in each case. For the squashed 7-sphere, using \eqref{eq:tracecanonicaltangentsphere} we obtain the solution of the heterotic G$_2$ system if and only if we set
\begin{equation}
    s=\frac{1}{\sqrt{2}}, \qquad \alpha'=\frac{2}{c-4}.
\end{equation}
Note the vector bundle representation has to be chosen such that $c>4$ in order to have a solution, which discards the SU(2) representations on $\C$ and $\C\oplus\C \, $.

For the squashed Aloff-Wallach space, using \eqref{eq:tracecanonicaltangentAloffWallach} the solution of the heterotic G$_2$ system is obtained if and only if we set
\begin{equation}
    s=\frac{1}{\sqrt{2}}, \qquad \alpha'=\frac{2}{3(q-4)}.
\end{equation}
Note that the representation has to be such that $q>4$, which rules out several low-dimensional representations.

Even though these solutions are isolated, they present the interesting feature that solutions with arbitrary small string parameter $\alpha'$ can be found by choosing a gauge bundle representation of sufficiently large dimension.

\subsubsection{Clarke-Oliveira connection on the tangent bundle}
\label{subsec:canCO}

Using \eqref{eq:traceclarkeoliveiratangent} we have two isolated solutions for each representation, both of them with the same value of $\alpha'$. One is always obtained for $s=\frac{1}{\sqrt{2}}$, where the Clarke-Oliveira connection reduces to the trivial flat connection on the tangent bundle. For the squashed 7-sphere the solutions are
\begin{equation}
    s=\frac{1}{\sqrt{2}} \ \ \text{or} \ \ s=\frac{\sqrt{12+c}}{2\sqrt{6}} \, , \qquad \alpha'=\frac{2}{c} \, ,
\end{equation}
whereas for the squashed Aloff-Wallach space we have
\begin{equation}
    s=\frac{1}{\sqrt{2}} \ \ \text{or} \ \ s=\frac{\sqrt{4+q}}{2\sqrt{2}}, \qquad \alpha'=\frac{2}{3 \, q} \, .
\end{equation}
In both cases all bundle representations are allowed. The string parameter $\alpha'$ can be made arbitrarily small by choosing a bundle representation of arbitrary large dimension.

\subsubsection{One-parameter family of connections on the tangent bundle}
\label{subsec:canfam}

The contribution of the one-parameter family to the heterotic Bianchi identity can be found in \eqref{eq:traceofgeneralg2connection} for the squashed 7-sphere or in \eqref{eq:traceofgeneralg2connectionAloffWallach} for the squashed Aloff-Wallach space. The solutions we find follow three different behaviours depending on the value of $c$ or $q$. The range where the solution is defined is controlled by the discriminant of the cubic equation for $a$, whose coefficients can be found in \cref{sec:cubiceqs}.

\begin{figure}[h]
\centering
\includegraphics[scale=0.7]{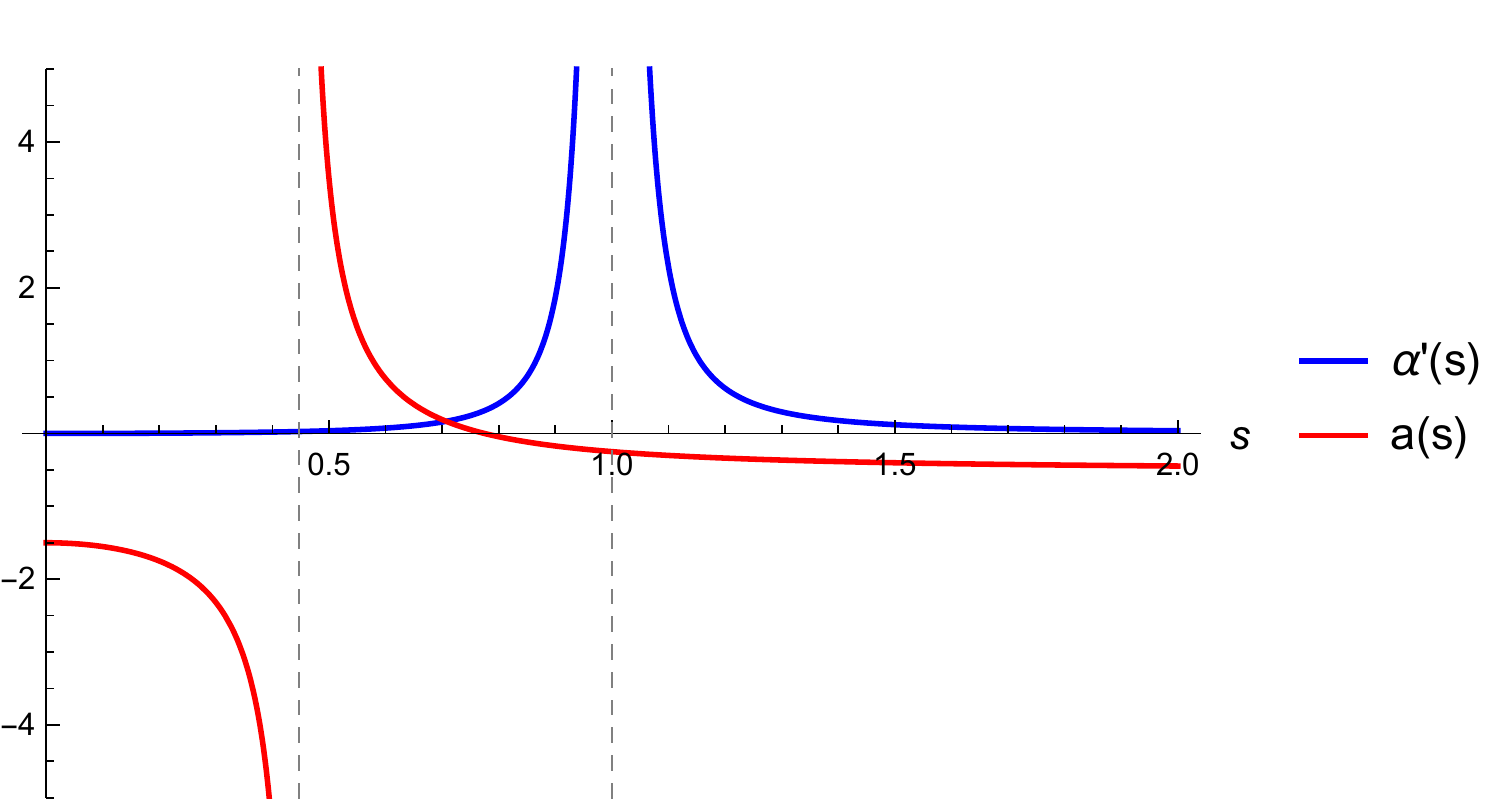}
\caption{Values of $\alpha'(s)$ and $a(s)$ in terms of $s$ for the squashed 7-sphere and the squashed Aloff-Wallach space solution \eqref{eq:easysolution}. The instanton in the tangent bundle belongs to the one-parameter family, with parameter $a(s)$. The instanton in the gauge bundle is the canonical connection in a representation with $c=4$. The dashed grey lines indicate the values of $s={\scriptstyle{1/\sqrt{5}}}\,, \, 1\,$, where no solutions exist.}
\label{fig:easysolution}
\end{figure}

The first case corresponds to representations with $c=4$ for the squashed 7-sphere or $q=4$ for the squashed Aloff-Wallach space\footnote{Note these representations include in particular the tangent bundle representation of the canonical connection. Since that particular representation is part of the one-parameter family---see \cref{foot:canisinfamily}---these solutions also appear when the one-parameter family is chosen in the vector bundle.}, in this case the discriminant vanishes identically and we obtain a unique solution for all $s$ except for the nearly parallel $s=\frac{1}{\sqrt{5}}$ and round $s=1$ values. The solution takes the same form for both the squashed 7-sphere and the squashed Aloff-Wallach space
\begin{equation}
\label{eq:easysolution}
    a(s)=-\frac{5 \, s^2-3}{10 \, s^2-2}, \qquad \alpha'(s)=\frac{s^2}{12(s^2-1)^2}.
\end{equation}
We show this solution in \cref{fig:easysolution}. Note the string parameter tends to 0 when $s\rightarrow 0$ and when $s\rightarrow\infty$, whereas it blows up whenever $s\rightarrow 1$. This means that the most interesting solutions from the point of view of physics are away from the round metric $s=1$.

For other values of $c$ or $q$ we find three different solutions which are well-defined only in certain ranges. We summarize them in \cref{tab:rangescanpar}, where we have introduced the quantity $s_1\in\left( \frac{1}{\sqrt{2}},1 \right)$ given implicitly for the squashed 7-sphere or the squashed Aloff-Wallach space by
\begin{equation}
    f(s_1)=c \qquad \text{or} \qquad g(s_1)=q \, ,
\end{equation}
respectively, where $f(s)$ and $g(s)$ were defined in \eqref{eq:fandg}.

\begin{table}[h]
{
\centering
\begin{tabular}{|c|c||c|c|c|c|}
\hline 
\multicolumn{2}{|c||}{Range of $s$} &
$\left(0,\frac{1}{\sqrt{5}}\right)$ \rule{0pt}{3ex} &
$\left(\frac{1}{\sqrt{5}},\frac{1}{\sqrt{2}}\right]$ &
$\left[\frac{1}{\sqrt{2}},s_1\right]$ &
$\left[s_1,\infty\right)$\\[1ex]
\hline
\hline
\multirow{3}{*}{$c>4$ or $q>4$} 
& Solution 1 & \checkmark &\checkmark &\checkmark &\checkmark \\ \cline{2-6}
& Solution 2 & & &\checkmark & \\ \cline{2-6}
& Solution 3 & & &\checkmark & \\
\hline
\multirow{3}{*}{$c<4$ or $q<4$} 
& Solution 1 &\checkmark & &\checkmark &\checkmark \\ \cline{2-6}
& Solution 2 & & & &\checkmark \\ \cline{2-6}
& Solution 3 & &\checkmark & &\checkmark \\
\hline
\end{tabular}
\caption{Ranges of solutions obtained for squashed homogeneous 7-dimensional 3-Sasakian manifolds with one-parameter family of connections in the tangent bundle and canonical connection in the vector bundle.
}
\label{tab:rangescanpar}
}
\end{table}

\begin{figure}[h]
\centering
\includegraphics[scale=0.7]{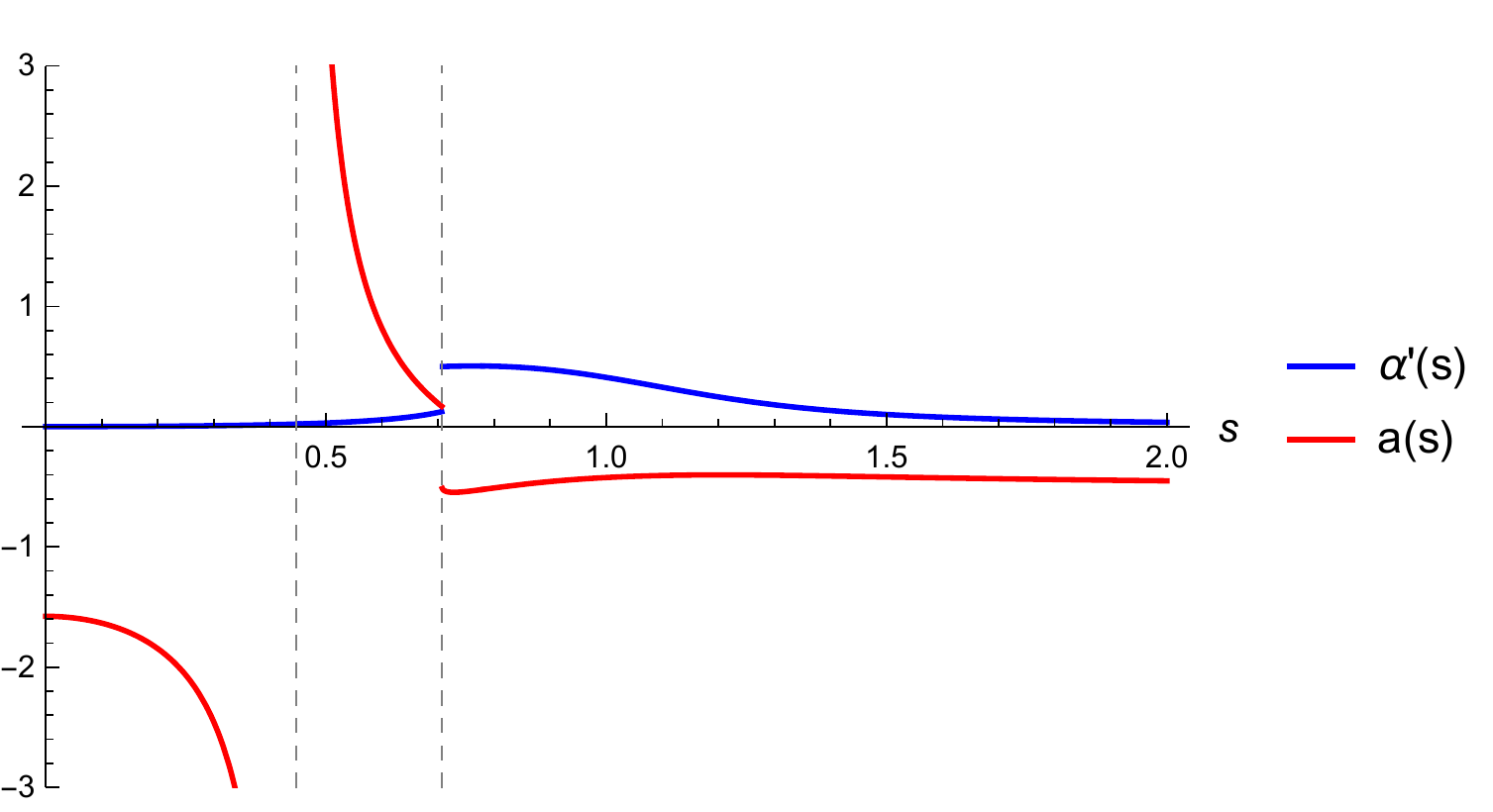}
\caption{Values of $\alpha'(s)$ and $a(s)$ in terms of $s$ for solution 1 in the squashed 7-sphere case. The instanton in the tangent bundle belongs to the one-parameter family, with parameter $a(s)$. The instanton in the gauge bundle is the canonical connection in a representation with $c=8$. The dashed grey lines indicate the values of $s={\scriptstyle{1/\sqrt{5}}}\,, \,{\scriptstyle{1/\sqrt{2}}}\,$. At $s={\scriptstyle{1/\sqrt{5}}}$ no solution exists whereas at $s={\scriptstyle{1/\sqrt{2}}}\,$ there is a discontinuity.}
\label{fig:oneparcanc8}
\end{figure}

Solution 1 for $c>4$ or $q>4$ is well defined for all $s$ except for $s=\frac{1}{\sqrt{5}}$. It is continuous at $s=s_1$, whereas at $s=\frac{1}{\sqrt{2}}$ there is a discrete jump: the left and right limits correspond to two different valid solutions
\begin{equation}
\label{eq:pairofsols1sqrt2}
    \left(a,\alpha'\right)=\left(\frac{1}{6} \, ,\frac{2}{z+8}\right), \qquad \left(a,\alpha'\right)=\left(- \, \frac{1}{2} \, ,\frac{2}{z-4}\right),
\end{equation}
where $z$ represents either $c$ or $q$. Similarly to the $c,q=4$ case, we have $\alpha'\rightarrow 0$ for both $s\rightarrow 0$ and $s\rightarrow\infty$. Nevertheless, $\alpha'$ does not blow up for any value of $s$. This provides a large number of solutions with a small value of the string parameter $\alpha'$. We show this solution for the squashed 7-sphere with the choice $c=8$ in \cref{fig:oneparcanc8}.
As for solutions 2 and 3 in the case $c>4$ or $q>4$, they are only defined in a very small range and they remain very close to the $s=\frac{1}{\sqrt{2}}$ solutions described in \eqref{eq:pairofsols1sqrt2}.

\begin{figure}[h]
\centering
\includegraphics[scale=0.7]{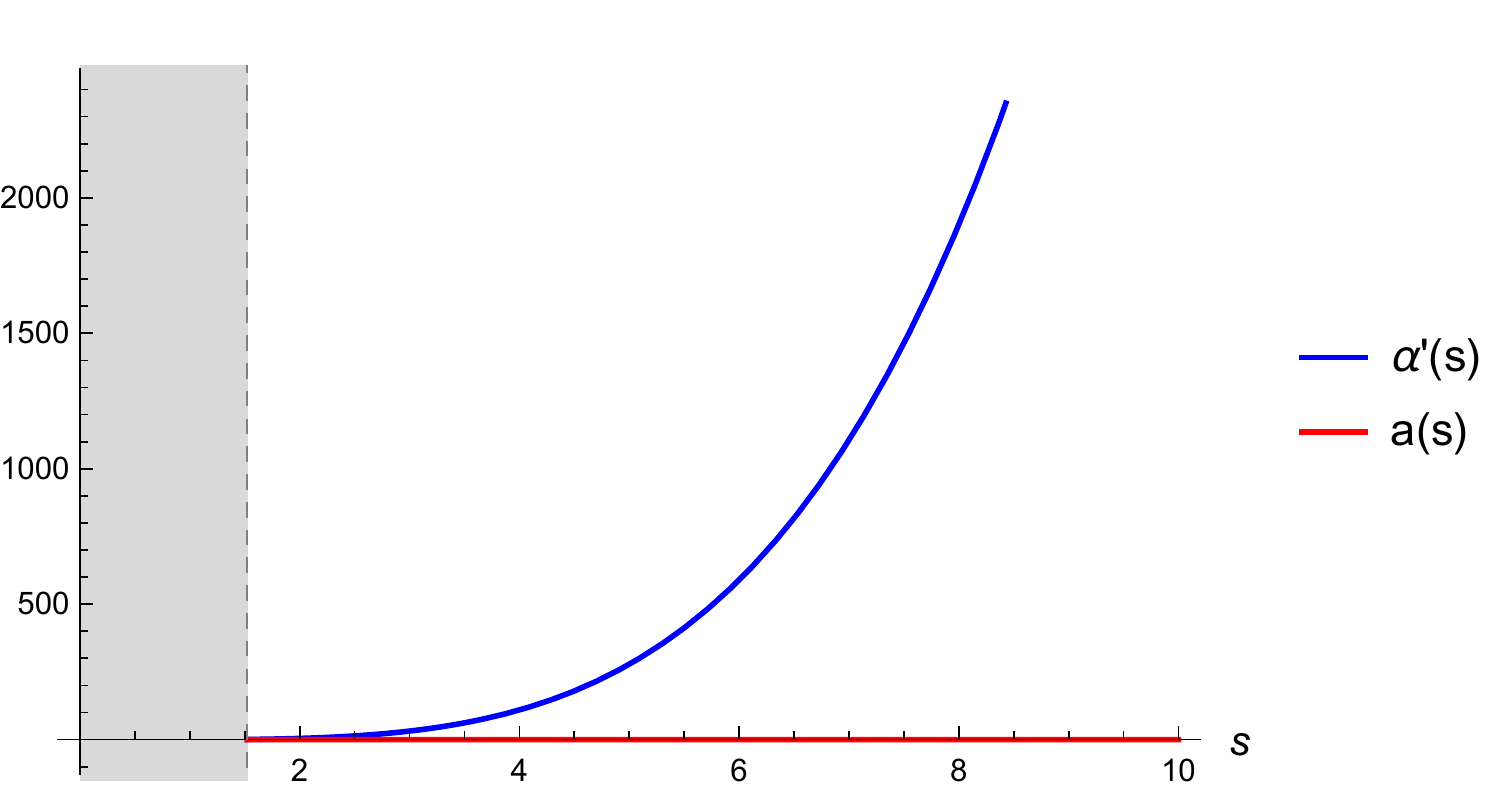}
\caption{Values of $\alpha'(s)$ and $a(s)$ in terms of $s$ for solution 2 in the squashed Aloff-Wallach case. The instanton in the tangent bundle belongs to the one-parameter family, with parameter $a(s)$. The instanton in the gauge bundle is the canonical connection in a representation with $q=1$. The grey region indicates values of $s<s_1$ where no solutions exist.}
\label{fig:oneparcanq1}
\end{figure}

When $c<4$ or $q<4$, solution 1 now presents a discrete jump at $s=s_1$. This solution still presents the physically interesting behaviour $\alpha'\rightarrow 0$ when $s\rightarrow 0$ or $s\rightarrow\infty$. On the other hand, solutions 2 and 3 satisfy $\alpha'\rightarrow\infty$ as $s\rightarrow\infty$. Even though these solutions are perfectly valid from a mathematical point of view, they are less attractive for physical purposes---we should think of $\alpha'$ as a small perturbative parameter, a property which is clearly not satisfied in this case. We illustrate this behaviour for solution 2 for the squashed Aloff-Wallach space case with $q=1$ in \cref{fig:oneparcanq1}.

\subsection{Clarke-Oliveira connection on the vector bundle}

In order to use the Clarke-Oliveira connection on the vector bundle we need to make a choice of representation for the SU(2) principal bundle as explained in \cref{sec:traceclarkeoliveira}. Take an arbitrary $\mathfrak{su}(2)$ representation, which is given as a direct sum of $k$ irreducible representations with Dynkin labels $m_1 \, ,\dots,m_k$. Let us denote for simplicity
\begin{equation}
    c=\left( c(m_1)+\cdots+c(m_k) \right),
\end{equation}
note $c$ can take any natural even value. Then the contribution of the Clarke-Oliveira connection to the heterotic Bianchi identity \eqref{eq:traceclarkeoliveira} is given by
\begin{equation}
\label{eq:abridgedtraceclarkeoliveira}
    \tr( F\wedge F)=-8 \, (1-2 \, s^2)^2 \, c\left( 3*_sF_1-2*_sF_2 \right),
\end{equation}
We study the available solutions depending on the choice of tangent bundle instanton.

\subsubsection{Canonical connection on the tangent bundle}

This particular choice of connections contributes with a negative coefficient to the $*_sF_1$ term in the Bianchi identity, as can be seen from \eqref{eq:tracecanonicaltangentsphere} or \eqref{eq:tracecanonicaltangentAloffWallach} and \eqref{eq:abridgedtraceclarkeoliveira}. On the other hand, the coefficient of $\dd H$ is positive, see \eqref{eq:exteriorderivativeflux}. Therefore, there are no solutions with positive string parameter $\alpha'$.

\subsubsection{Clarke-Oliveira connection on the tangent bundle}

In this case the $*_sF_1$ and $*_sF_2$ terms corresponding to the connections in the heterotic Bianchi identity are proportional to each other, and they can not be equal to the $\dd H$ contribution while keeping $\alpha'>0 \, $.

\subsubsection{One-parameter family of connections on the tangent bundle}
\label{subsec:COfam}

For each value of $c$ we find three different solutions. The domain of these solutions depends on the value of $c$ itself and on the chosen homogeneous space, so we describe the solutions for the squashed 7-sphere and the squashed Aloff-Wallach space separately.

For the squashed 7-sphere, the contribution of the one-parameter family to the heterotic Bianchi identity can be found in \eqref{eq:traceofgeneralg2connection}. The domain of solutions is determined by the discriminant of the cubic equation for $a$ (see \cref{sec:cubiceqs}) and the positive sign of $\alpha'$, as explained around equation \eqref{eq:fandg}. This is controlled by three quantities which depend on the chosen value of $c$. We denote them by $s_2$, $s_3$ and $s_4$. We define $s_2$ and $s_4$ as the positive roots of the equation
\begin{equation}
\label{eq:equationfor-fs}
    -f(s)=c \, ,
\end{equation}
where $f(s)$ was defined in \eqref{eq:fandg}, $\frac{1}{\sqrt{2}}<s_4$ and $s_2$ is only present in the case $c>10 \, $, determined by $s_2<\frac{1}{\sqrt{2}}$. We define $s_3$ as the positive real root of the polynomial
\begin{equation}
    -\left( c \, (1-2 \, s^2) \right)^3-48 \, s^2(1-s^2)\left( c \, (1-2 \, s^2) \right)^2+48 \, (1-4 \, s^2) \,  c \, (1-2 \, s^2) +128 \, .
\end{equation}
We show the ranges of the solutions in \cref{tab:rangesclarpar1}.

\begin{table}[h]
{
\centering
\begin{tabular}{|c|c||c|c|c|c|c|c|}
\hline 
\multicolumn{2}{|c||}{Range of $s$} &
$\left(0,\frac{1}{\sqrt{5}}\right)$ \rule{0pt}{3ex} &
$\left(\frac{1}{\sqrt{5}},s_2\right]$ &
$\left(s_2,s_3\right)$ &
$\left[s_3,\frac{1}{\sqrt{2}}\right]$ &
$\left[\frac{1}{\sqrt{2}},s_4\right]$ &
$\left[s_4,\infty\right)$\\[1ex]
\hline
\hline
\multirow{3}{*}{$c<8$} 
& Solution 1 & \checkmark & & & &\checkmark &\checkmark \\ \cline{2-8}
& Solution 2 & & & & & &\checkmark \\ \cline{2-8}
& Solution 3 & &\checkmark &\checkmark &\checkmark & &\checkmark \\
\hline
\multirow{3}{*}{$10<c<40$} 
& Solution 1 & & & & &\checkmark &\checkmark \\ \cline{2-8}
& Solution 2 & & &\checkmark & & &\checkmark \\ \cline{2-8}
& Solution 3 & & &\checkmark &\checkmark & &\checkmark \\
\hline
\multirow{3}{*}{$40\le c$} 
& Solution 1 & & & & &\checkmark &\checkmark \\ \cline{2-8}
& Solution 2 & & & & & &\checkmark \\ \cline{2-8}
& Solution 3 & & & &\checkmark & &\checkmark \\
\hline
\end{tabular}
\caption{Ranges of solutions obtained for the squashed 7-sphere with one-parameter family of connections in the tangent bundle and Clarke-Oliveira connection in the vector bundle.
}
\label{tab:rangesclarpar1}
}
\end{table}

Only two cases are absent in the table. For the case $c=10$ the equation \eqref{eq:equationfor-fs} has a third positive root that we denote $\tilde{s_2}$, with $\tilde{s_2}<s_2<s_4 \, $. The behaviour in this case is as in $10<c<40$ with the addition that solutions 1 and 2 are also well-defined in the interval $(0,\tilde{s_2}] \, $.

\begin{figure}[h]
\centering
\includegraphics[scale=0.7]{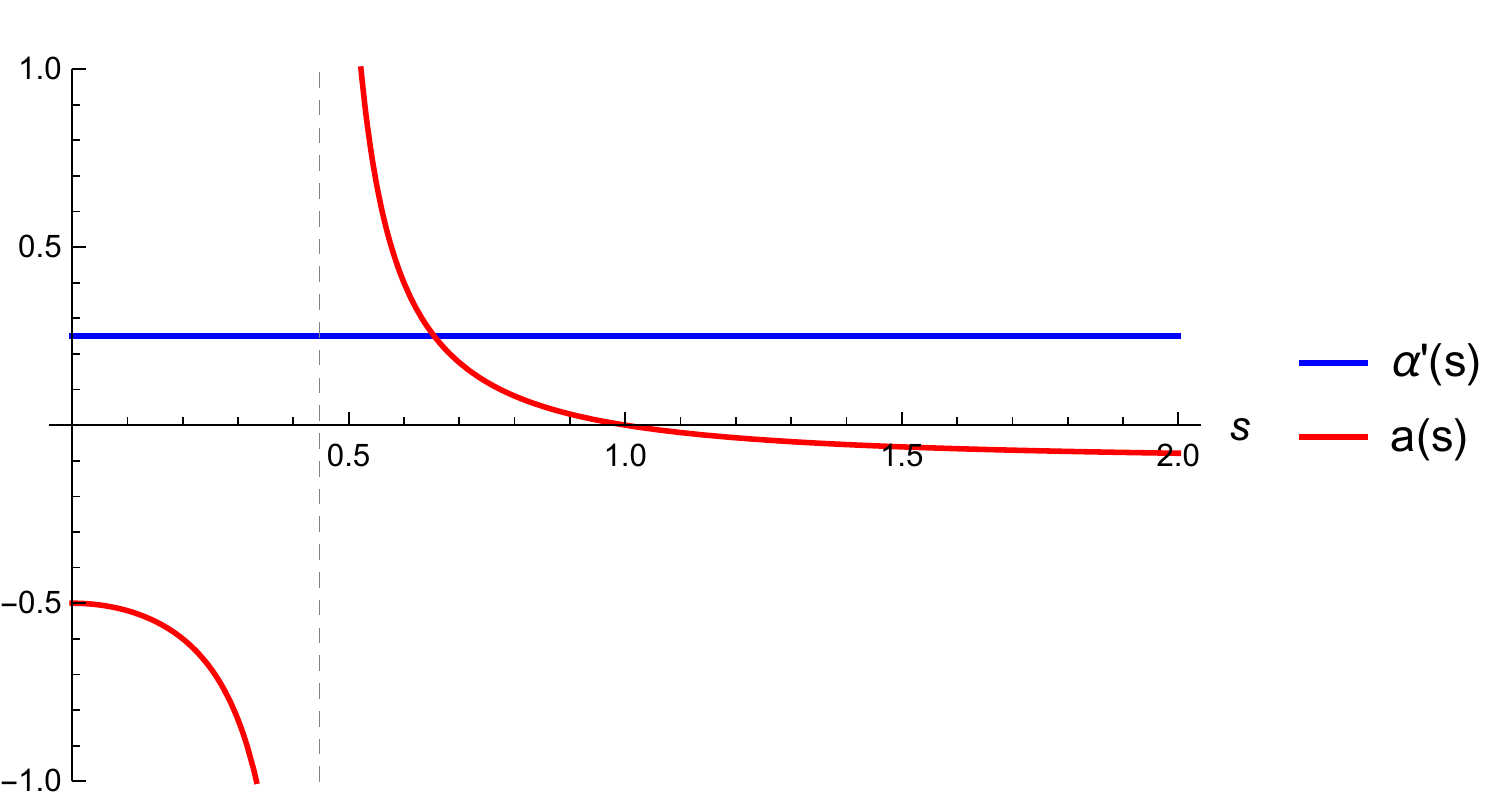}
\caption{Values of $\alpha'(s)$ and $a(s)$ in terms of $s$ for solution 3 in the squashed 7-sphere case. The instanton in the tangent bundle belongs to the one-parameter family, with parameter $a(s)$. The instanton in the gauge bundle is the Clarke-Oliveira connection in a representation with $c=8$. The dashed grey line indicates the value $s=\scriptstyle{1/\sqrt{5}}$ where no solution exists.}
\label{fig:oneparCOc8}
\end{figure}

For the case $c=8$ something remarkable happens. Solution 1 has the same range as for $c<8$ and solution 2 incorporates the interval $\left(\frac{1}{\sqrt{5}},s_2\right]$ to the range we find for $c<8$. Solution 3, on the other hand, is very special: it is defined everywhere except for the nearly-parallel value $s=\frac{1}{\sqrt{5}}$ and the string parameter is fixed to a constant value $\alpha'=\frac{1}{4} \, $. We show this solution in \cref{fig:oneparCOc8}. This is, to the best of our knowledge, the first example of a family of solutions of the heterotic G$_2$ system with a fixed value of $\alpha'<1$. Keeping the string parameter $\alpha'$ fixed, we can deform the G$_2$-structure and the instanton connection in the tangent bundle from a fixed value of $s$. We comment further on this in \cref{sec:conclusions}.

Now we discuss \cref{tab:rangesclarpar1} in more generality. For all values of $c$, solution 1 is discontinuous at $s=s_4$. The value of $\alpha'$ remains small for all the range of $s$ and in fact $\alpha'\rightarrow 0$ when $s\rightarrow\infty$. This means for every choice of $c$ we can obtain solutions with arbitrary small string parameter $\alpha'$.

\begin{figure}[h]
\centering
\includegraphics[scale=0.7]{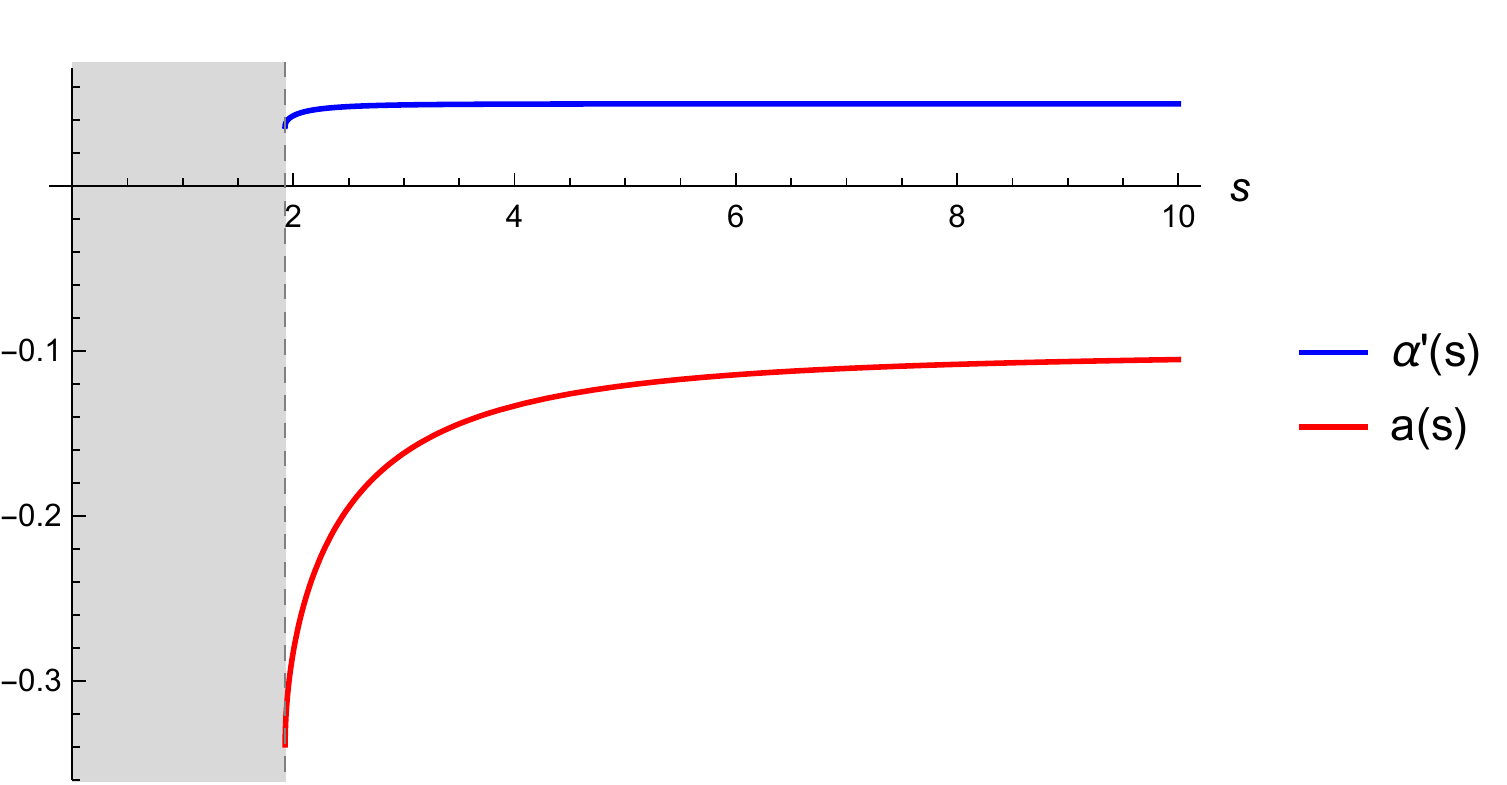}
\caption{Values of $\alpha'(s)$ and $a(s)$ in terms of $s$ for solution 2 in the squashed 7-sphere case. The instanton in the tangent bundle belongs to the one-parameter family, with parameter $a(s)$. The instanton in the gauge bundle is the Clarke-Oliveira connection in a representation with $c=40$. The grey region indicates values of $s<s_4$ where no solutions exist.}
\label{fig:oneparCOc20}
\end{figure}

Solutions 2 and 3 present a slightly different behaviour. First of all, they do not present discontinuities in their domains. Within the interval $\left(\frac{1}{\sqrt{5}},\frac{1}{\sqrt{2}}\right)$, the solutions are typically defined over a very small range and the value of $\alpha'$ blows up. Therefore, that part of the solutions is less amenable to an interesting physical interpretation. On the other hand, we always find the limiting behaviour $\alpha'\rightarrow \frac{2}{c}$ when $s\rightarrow\infty$. In fact, the solutions soon stabilize in values close to $\alpha'=\frac{2}{c} \, $. This means we find solutions with an almost constant value of $\alpha'$ which can be made arbitrarily small by choosing different representations of the gauge bundle. We illustrate this for the choice $c=40$ in \cref{fig:oneparCOc20}.

\bigskip

For the squashed Aloff-Wallach space, the contribution of the one-parameter family to the heterotic Bianchi identity can be found in \eqref{eq:traceofgeneralg2connectionAloffWallach}. The domain of the solutions is simpler and we just distinguish two cases. These are shown in \cref{tab:rangesclarpar2}: all solutions are well defined beyond a certain value $s_4$, which is given by the highest root of the equation
\begin{equation}
\label{eq:equationfor-3gs}
    -3g(s)=c \, ,
\end{equation}
for $g(s)$ as defined in \eqref{eq:fandg}. When $c=2$ we have an additional interval in the domain determined by the smallest positive root of equation \eqref{eq:equationfor-3gs}, which we denote by $s_2$.

\begin{table}[h]
{
\centering
\begin{tabular}{|c|c||c|c|c|c|}
\hline 
\multicolumn{2}{|c||}{Range of $s$} &
$\left(0,s_2\right]$ \rule{0pt}{3ex} &
$\left(s_2,\frac{1}{\sqrt{2}}\right]$ &
$\left[\frac{1}{\sqrt{2}},s_4\right]$ &
$\left[s_4,\infty\right)$\\[1ex]
\hline
\hline
\multirow{3}{*}{$c=2$} 
& Solution 1 & \checkmark & &\checkmark &\checkmark \\ \cline{2-6}
& Solution 2 & \checkmark & & &\checkmark \\ \cline{2-6}
& Solution 3 & & & &\checkmark \\
\hline
\multirow{3}{*}{$c>2$} 
& Solution 1 & & &\checkmark &\checkmark \\ \cline{2-6}
& Solution 2 & & & &\checkmark \\ \cline{2-6}
& Solution 3 & & & &\checkmark \\
\hline
\end{tabular}
\caption{Ranges of solutions obtained for the squashed Aloff-Wallach space with one-parameter family of connections in the tangent bundle and Clarke-Oliveira connection in the vector bundle.
}
\label{tab:rangesclarpar2}
}
\end{table}

\begin{figure}[h]
\centering
\includegraphics[scale=0.7]{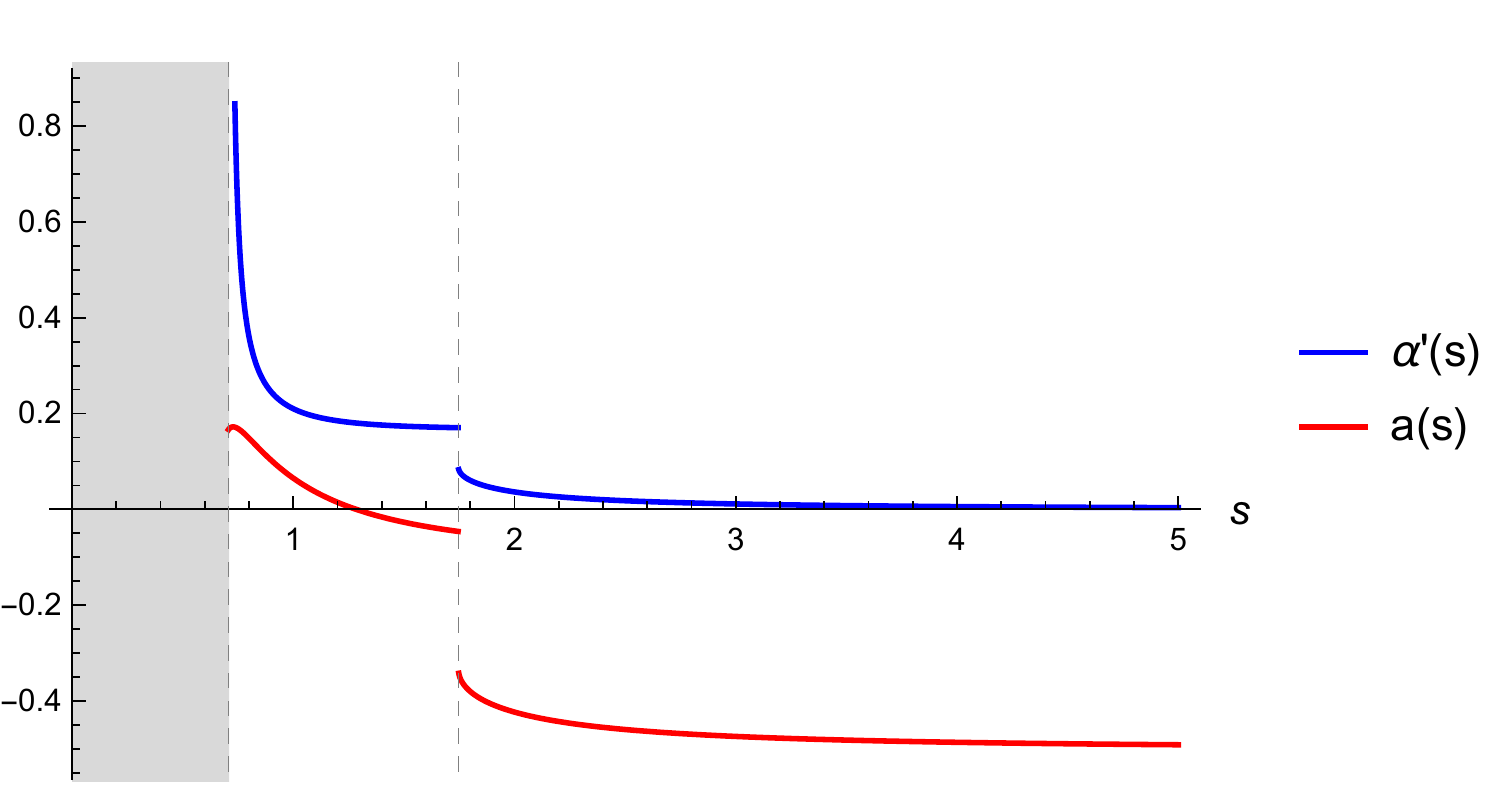}
\caption{Values of $\alpha'(s)$ and $a(s)$ in terms of $s$ for solution 1 in the squashed Aloff-Wallach case. The instanton in the tangent bundle belongs to the one-parameter family, with parameter $a(s)$. The instanton in the gauge bundle is the Clarke-Oliveira connection in a representation with $c=12$. The grey region indicates values of $s<{\scriptstyle{1/\sqrt{2}}}$ where no solutions exist, and the dashed line indicates a discontinuity of the solution at $s=s_4$.}
\label{fig:oneparCOc12}
\end{figure}

The behaviour of the solutions is very similar to the squashed 7-sphere case we have just described. For all values of $c$, solution 1 is discontinuous at $s=s_4$ and we have $\alpha'\rightarrow 0$ when $s\rightarrow\infty$. We present an illustrative example with $c=20$ in \cref{fig:oneparCOc12}. As for solutions 2 and 3, we find again the same behaviour $\alpha'\rightarrow \frac{2}{c}$ when $s\rightarrow\infty$ and solutions stay almost constant for the whole range of $s$.

\newpage

\subsection{One-parameter family of connections on the vector bundle}

For this last set of solutions we choose the tangent bundle $TY$ as the vector bundle $V$, and a connection from the one-parameter family of \cref{sec:tangentbundleinstantons} as the vector bundle G$_2$-instanton. As we will see, our solutions have different connections on each bundle. We denote the parameter of this family as $b$, and the contribution to the heterotic Bianchi identity is given by \eqref{eq:traceofgeneralg2connection} for the squashed 7-sphere
\begin{equation}
    \tr(F\wedge F)= -72 \,  s^2 \left( \kappa(b,s)^2 -\frac{4}{3 \, s^2} \right)*_sF_1 -12 \, s \ \kappa(b,s)^2\left(\kappa(b,s)-\frac{2}{s}\right)*_sF_2 \, .
\end{equation}
and \eqref{eq:traceofgeneralg2connectionAloffWallach} for the squashed Aloff-Wallach space
\begin{equation}
    \tr(F\wedge F)= -72 \,  s^2 \left( \kappa(b,s)^2 -\frac{4}{s^2} \right)*_sF_1 -12 \, s \ \kappa(b,s)^2\left(\kappa(b,s)-\frac{2}{s}\right)*_sF_2 \, .
\end{equation}
where $\kappa(b,s)=(1+10 \, b)s+(1-2 \, b)/s$. We find different solutions in terms of the choice of tangent bundle instanton.

\subsubsection{Canonical connection on the tangent bundle}

The equations for $*_sF_1$ and $*_sF_2$ in this case have a unique solution that is exactly as the one presented in \eqref{eq:easysolution} but with opposite $\alpha'$ sign. As a result, we can not impose $\alpha'>0$ and there are no solutions for this choice of instantons.

\subsubsection{Clarke-Oliveira connection on the tangent bundle}
\label{subsec:famCO}

We treat the squashed 7-sphere and the squashed Aloff-Wallach space separately. For each of them we have three solutions with domains given by the discriminant of the cubic equation for the parameter $b$ and by the condition $\alpha'>0$, as explained around equation \eqref{eq:fandg}. The coefficients of the cubic equation are listed in \cref{sec:cubiceqs}.

For the squashed 7-sphere, there are three relevant numbers $s_5$, $s_6$ and $s_7$ that we define as follows: $s_5$ and $s_7$ are the positive roots of the polynomial
\begin{equation}
    16s^6+6s^4-15s^2+2 \, ,
\end{equation}
with $s_5<s_7$. On the other hand, $s_6$ is the only positive root of the polynomial
\begin{equation}
    -36s^6+36s^4-15s^2+2 \, .
\end{equation}
We have detailed the range of the solutions in \cref{tab:rangesparclar1}. In addition to the ranges depicted, solutions 1 and 2 can be extended to $s=\sqrt{\frac{2}{3}} \, $.

\begin{table}[h]
{
\centering
\begin{tabular}{|c||c|c|c|c|c|c|c|}
\hline 
Range of $s$ &
$\left(0,s_5\right)$ \rule{0pt}{3.5ex} &
$\left[s_5,\frac{1}{\sqrt{5}}\right)$ &
$\left(\frac{1}{\sqrt{5}},s_6\right]$ &
$\left(s_6,\frac{1}{\sqrt{2}}\right]$ &
$\left(\frac{1}{\sqrt{2}},\sqrt{\frac{2}{3}}\right)$ &
$\left(\sqrt{\frac{2}{3}},s_7\right]$ &
$\left(s_7,\infty\right)$\\[1ex]
\hline
\hline
 Solution 1 & \checkmark & & \checkmark & \checkmark & &\checkmark & \checkmark \\ \hline
 Solution 2 & & & & \checkmark & &\checkmark & \\ \hline
 Solution 3 & &\checkmark & & & &\checkmark & \\
\hline
\end{tabular}
\caption{Ranges of solutions obtained for the squashed 7-sphere with Clarke-Oliveira connection in the tangent bundle and one-parameter family of connections in the vector bundle.
}
\label{tab:rangesparclar1}
}
\end{table}

\begin{figure}[h]
\centering
\includegraphics[scale=0.7]{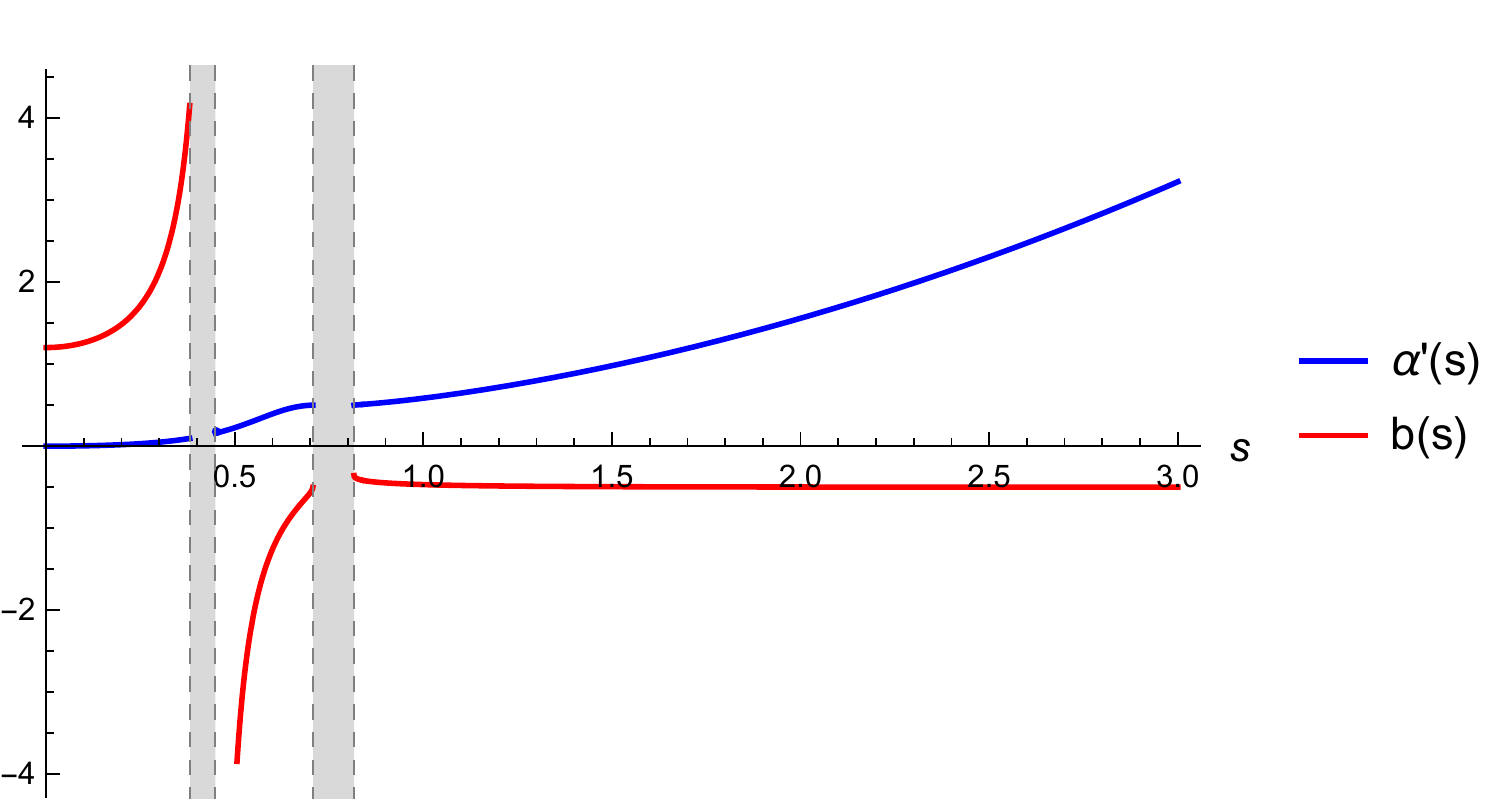}
\caption{Values of $\alpha'(s)$ and $b(s)$ in terms of $s$ for solution 1 in the squashed 7-sphere case. The instanton in the tangent bundle is the Clarke-Oliveira connection in the tangent bundle representation. The instanton in the gauge bundle belongs to the one-parameter family, with parameter $b(s)$. The grey regions indicate values of $s_5<s<{\scriptstyle{1/\sqrt{5}}}$ and ${\scriptstyle{1/\sqrt{2}}}<s<{\scriptstyle{\sqrt{2}/\sqrt{3}}}$ where no solutions exist.}
\label{fig:COonepar}
\end{figure}

Note that solution 1 is the only one defined for $s\rightarrow\infty$, in this limit we have $\alpha'\rightarrow\infty$. It is also the only solution defined for $s\rightarrow 0$ and in this case we find $\alpha'\rightarrow 0$. It is continuous at $s_6$ and $s_7$. These features can be observed in \cref{fig:COonepar}. Solutions 2 and 3 are defined in two small intervals and the string parameter $\alpha'$ blows up for one of them, making the solutions less interesting.

\bigskip

For the squashed Aloff-Wallach space, the ranges are simpler. We introduce the number
\begin{equation}
    s_8=\frac{\sqrt{3\sqrt{33}-11}}{4} \, ,
\end{equation}
and we list the domain of each solution in \cref{tab:rangesparclar2}. In addition to the ranges depicted, solutions 2 and 3 can be extended to $s=\frac{1}{\sqrt{2}}$ preserving continuity from the left.

\begin{table}[h]
{
\centering
\begin{tabular}{|c||c|c|c|c|}
\hline 
Range of $s$ &
$\left(0,\frac{1}{\sqrt{5}}\right)$ &
$\left(\frac{1}{\sqrt{5}},s_8\right]$ &
$\left[s_8,\frac{1}{\sqrt{2}}\right)$ \rule{0pt}{3ex} &
$\left(\frac{1}{\sqrt{2}},\infty\right)$\\[1ex]
\hline
\hline
 Solution 1 & \checkmark & \checkmark &\checkmark & \checkmark \\ \hline
 Solution 2 & & & \checkmark &  \\ \hline
 Solution 3 & & & \checkmark  & \\
\hline
\end{tabular}
\caption{Ranges of solutions obtained for the squashed Aloff-Wallach space with Clarke-Oliveira connection in the tangent bundle and one-parameter family of connections in the vector bundle.
}
\label{tab:rangesparclar2}
}
\end{table}

\smallskip

\begin{figure}[h]
\centering
\includegraphics[scale=0.7]{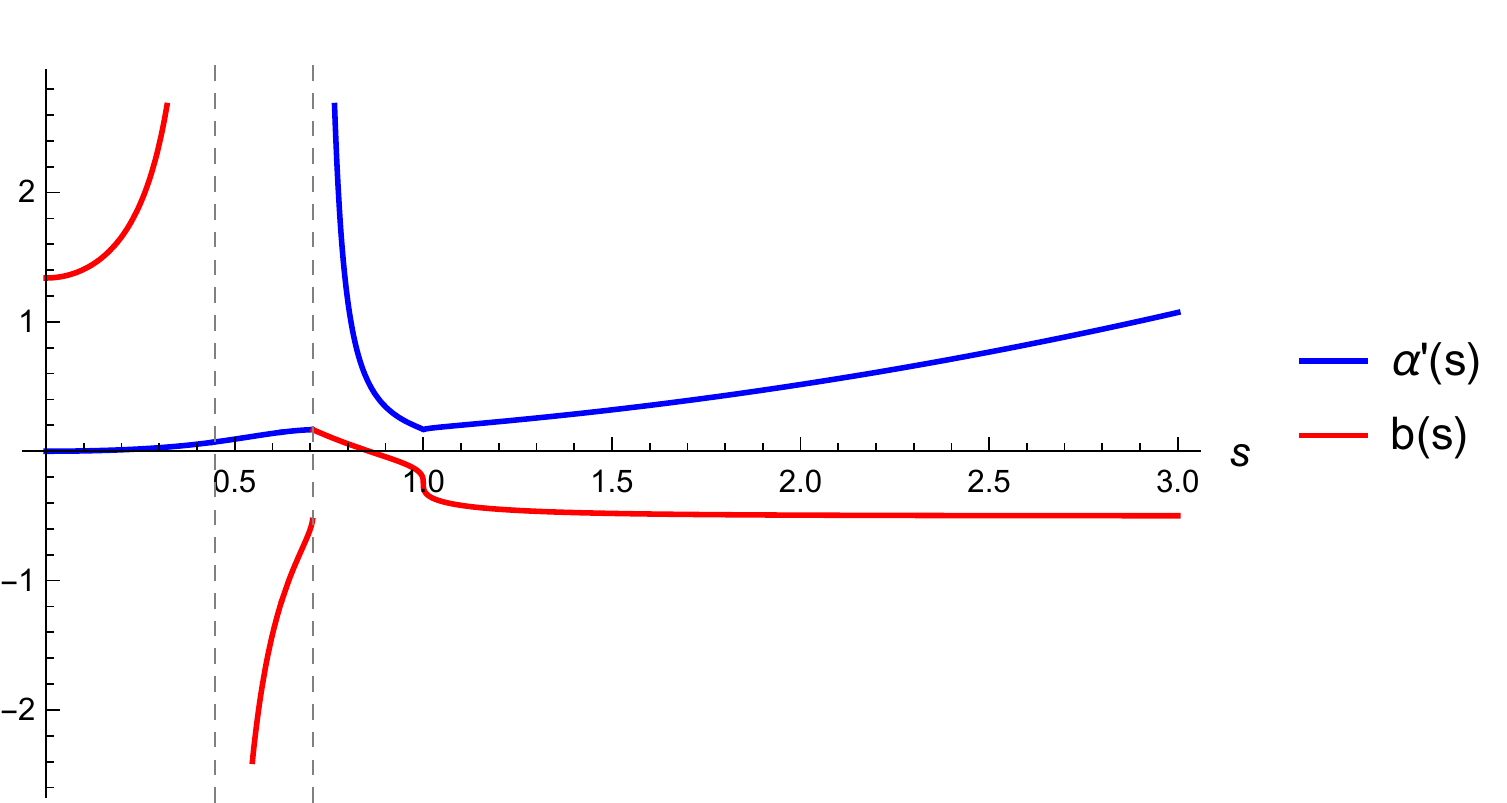}
\caption{Values of $\alpha'(s)$ and $b(s)$ in terms of $s$ for solution 1 in the squashed Aloff-Wallach case. The instanton in the tangent bundle is the Clarke-Oliveira connection in the tangent bundle representation. The instanton in the gauge bundle belongs to the one-parameter family, with parameter $b(s)$. The dashed grey lines indicate the values of $s={\scriptstyle{1/\sqrt{5}}}\,, \, {\scriptstyle{1/\sqrt{2}}}\,$, where no solutions exist.}
\label{fig:COoneparAW}
\end{figure}

Solution 1 is continuous at $s_8$ and discontinuous at $\frac{1}{\sqrt{5}}$ and $\frac{1}{\sqrt{2}}$, where no solutions exist. We find $\alpha'\rightarrow\infty$ as $s\rightarrow\infty$ and $\alpha'\rightarrow 0$ as $s\rightarrow 0$. We plot the solution in \cref{fig:COoneparAW}. Solutions 2 and 3 are again defined in a very small interval and $\alpha'$ blows up.

\subsubsection{One-parameter family of connections on the tangent bundle}
\label{subsec:famfam}

In this case we choose connections from the one-parameter family both for the vector bundle and the tangent bundle, and we denote the parameters by $b$ and $a$ respectively. Having two continuous parameters allows for a wider range of solutions, as we now explain.

Both instantons have the same contribution to the heterotic Bianchi identity but with opposite signs. This means the equations for $*_sF_1$ and $*_sF_2$ have the same coefficients for $a$ and $b$ up to the sign. The $*_sF_1$ equation is quadratic in $a$ and $b$ whereas the $*_sF_2$ equation is cubic in $a$ and $b$. We impose these equations together with the condition $\alpha'>0$.

\begin{figure}[h]
\centering
\includegraphics[scale=1]{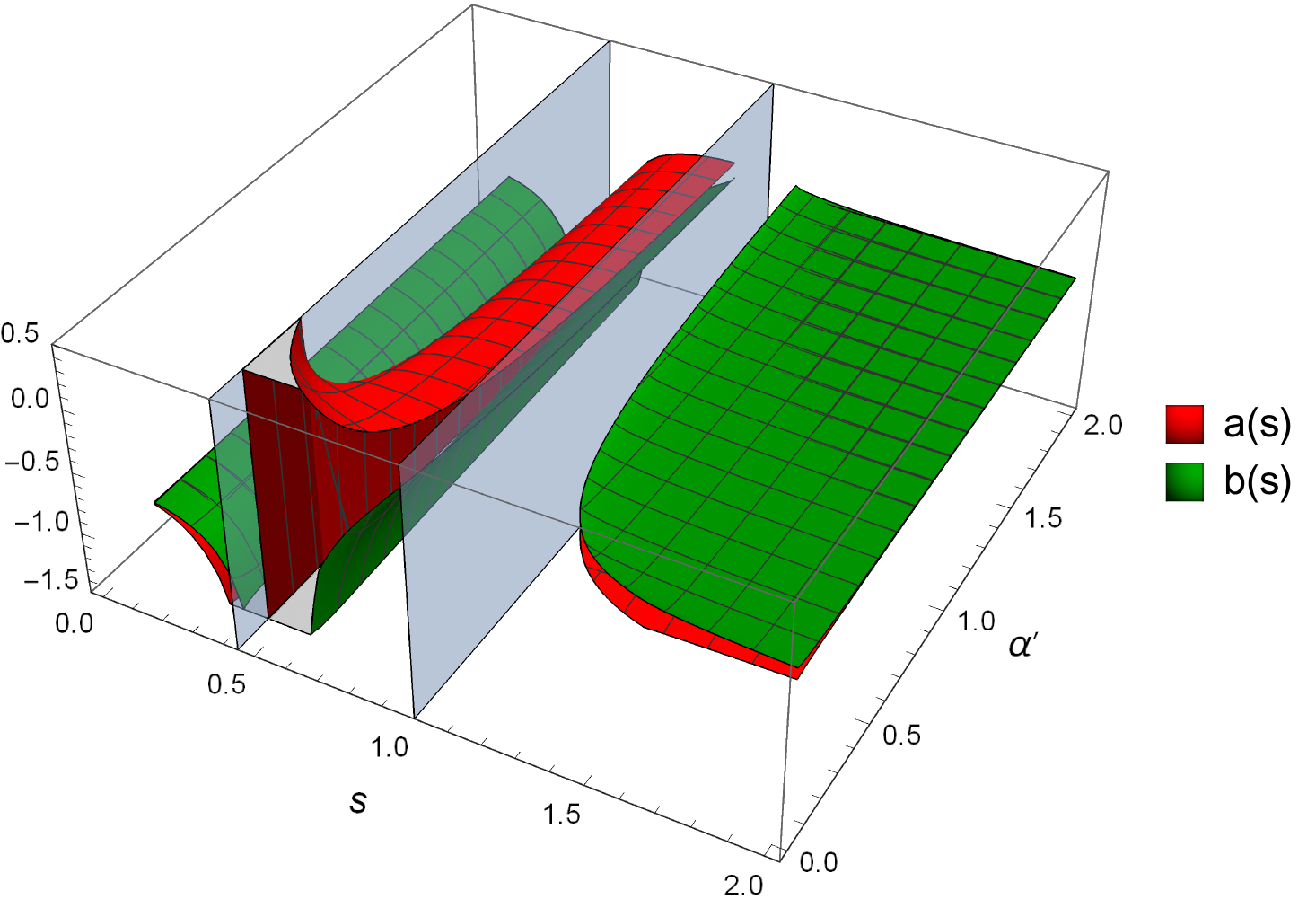}
\caption{First set of values of $a(s,\alpha')$ and $b(s,\alpha')$ in terms of $s$ and $\alpha'$. These provide solutions for both the squashed 7-sphere and the squashed Aloff-Wallach space. The instantons in both the tangent bundle and the gauge bundle belong to the one-parameter family, with parameters $a(s,\alpha')$ and $b(s,\alpha')$ respectively. The grey planes indicate values of $s={\scriptstyle{1/\sqrt{5}}}\,, \, 1\,$, where no solutions exist.}
\label{fig:FamFam1}
\end{figure}

\begin{figure}[h]
\centering
\includegraphics[scale=1]{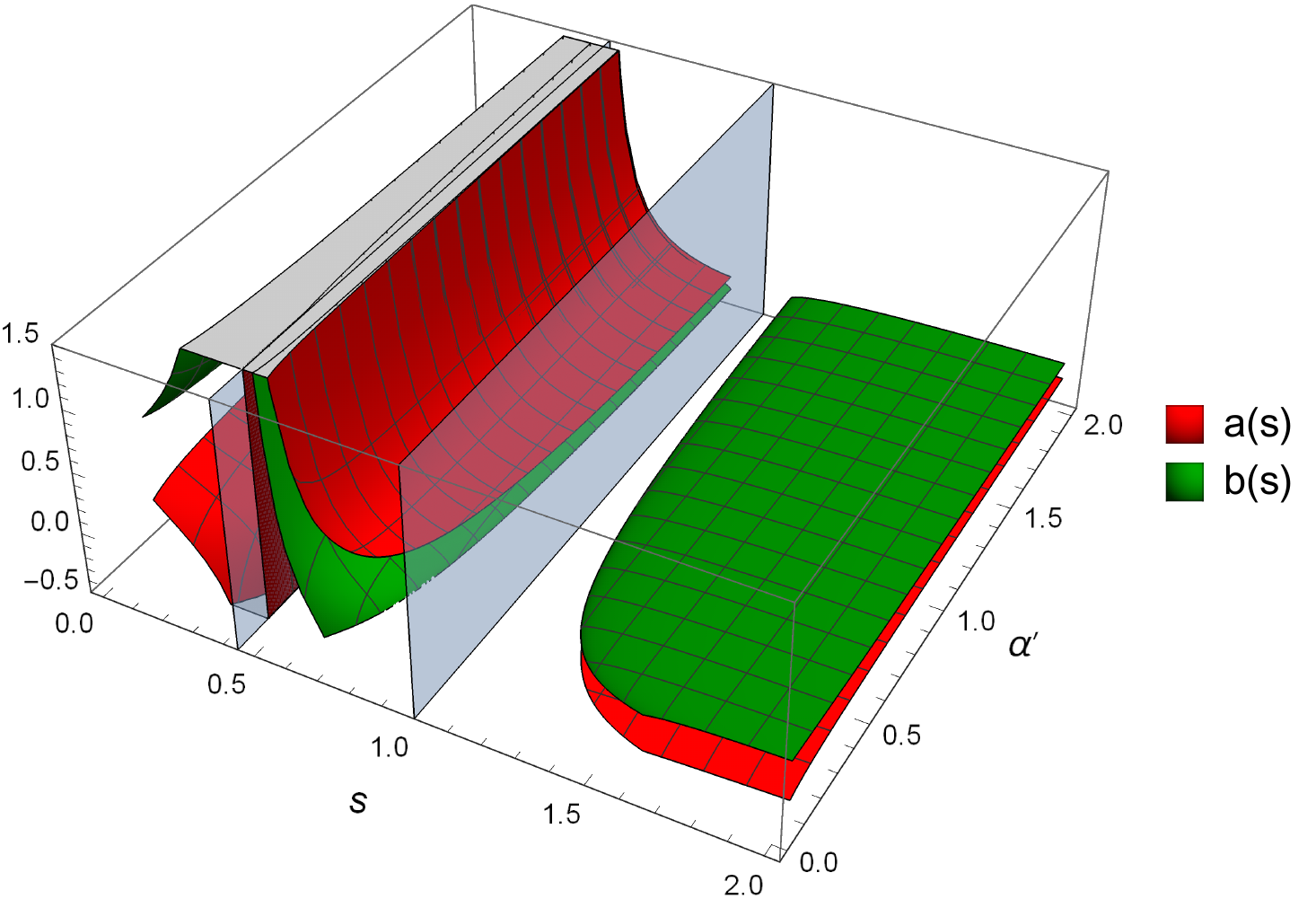}
\caption{Second set of values of $a(s,\alpha')$ and $b(s,\alpha')$ in terms of $s$ and $\alpha'$. These provide solutions for both the squashed 7-sphere and the squashed Aloff-Wallach space. The instantons in both the tangent bundle and the gauge bundle belong to the one-parameter family, with parameters $a(s,\alpha')$ and $b(s,\alpha')$ respectively. The grey planes indicate values of $s={\scriptstyle{1/\sqrt{5}}}\,, \, 1\,$, where no solutions exist.}
\label{fig:FamFam2}
\end{figure}

Both for the squashed 7-sphere and the squashed Aloff-Wallach space, we find that for every value of the squashing parameter $s$ except for $s=1$ and $s=\frac{1}{\sqrt{5}}$, and for every value of the string parameter $\alpha'$ such that
\begin{equation}
\label{eq:existsolcond}
    \alpha'> \frac{s^2}{12(s^2-1)^2} \, ,
\end{equation}
we have two sets of values for $a(s,\alpha')$ and $b(s,\alpha')$ that solve the heterotic Bianchi identity \eqref{eq:heteroticBianchiidentity} and provide solutions of the heterotic G$_2$ system. We present plots of the solutions in \cref{fig:FamFam1,fig:FamFam2}.

If we consider the limit value $\alpha'=\frac{s^2}{12(s^2-1)^2}$, we find a unique solution of the form
\begin{equation}
\label{eq:limitsolution}
    a(s)=-\frac{5 \, s^2-3}{10s^2-2} \, , \qquad b(s)=-\frac{s^2+1}{10 \, s^2-2}.
\end{equation}
Note this is precisely the same solution we found in \eqref{eq:easysolution}. This is because for this specific choice of parameter $b(s)$ the instanton coincides with the tangent bundle representation of the canonical connection, see \cref{foot:canisinfamily}.

The solutions we find present a very interesting profile. As is the case with other solutions, the parameters of the connections blow up at $s=\frac{1}{\sqrt{5}}$ reflecting the fact that the family of instantons collapses to a single connection. For $s=1$ it is not possible to obtain a solution with finite $\alpha'$, as is the case for the solution depicted in \cref{fig:easysolution}. In fact, a whole region around $s=1$ is excluded by the condition \eqref{eq:existsolcond}.

When $s\rightarrow 0$ or $s\rightarrow\infty$, the parameters of both connections tend to the same value independently of the value of $\alpha'$.\footnote{Note in these limits the connections tend to a configuration analogous to the standard embedding, where the curvatures from both connections cancel each other.} For the first set (\cref{fig:FamFam1}) we find $a,b\rightarrow-\frac{5}{6}$ as $s\rightarrow 0$, and $a,b\rightarrow-\frac{11}{30}$ as $s\rightarrow \infty$. For the second set (\cref{fig:FamFam2}) we find $a,b\rightarrow\frac{1}{2}$ as $s\rightarrow 0$, and $a,b\rightarrow-\frac{1}{10}$ as $s\rightarrow \infty$. Interestingly, for $s>1$ the parameters of the connections remain almost constant. Analogously, varying the string parameter $\alpha'$ has a very mild effect on the parameters.

One of the most interesting features of these solutions is that they can be regarded as finite deformations from a given solution. For a fixed value of $\alpha'$, the solutions describe a deformation with the squashing parameter $s$ as the deformation parameter. Together with the solution depicted in \cref{fig:oneparCOc8}, these are the first examples of finite deformations of the heterotic G$_2$ system. See \cite{delaOssa:2017pqy} for a description of the infinitesimal deformations.

\begin{figure}[h]
\centering
\includegraphics[scale=0.7]{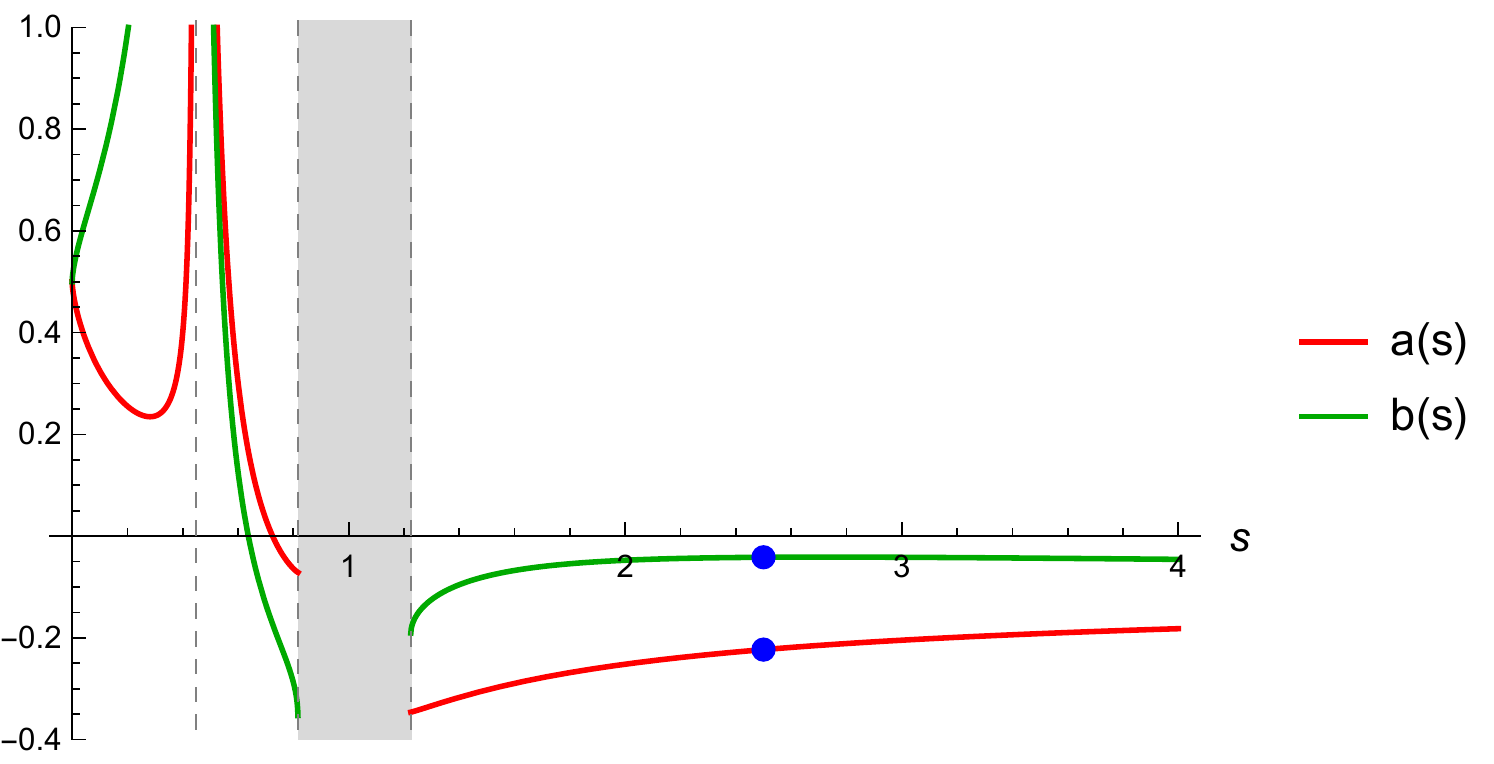}
\caption{One set of values of $a(s,\alpha')$ and $b(s,\alpha')$ in terms of $s$ and with $\alpha'=1/2$ fixed. These provide solutions for both the squashed 7-sphere and the squashed Aloff-Wallach space. The instantons in both the tangent bundle and the gauge bundle belong to the one-parameter family, with parameters $a(s,\alpha')$ and $b(s,\alpha')$ respectively. The blue dots refer to the starting point for the deformation $s={\scriptstyle{5/2}}$. The grey region indicates values of ${\scriptstyle{\sqrt{2}/\sqrt{3}}}<s<{\scriptstyle{\sqrt{3}/\sqrt{2}}}$ where no solutions exist, and the dashed line indicates the value $s={\scriptstyle{1/\sqrt{5}}}$ where no solution exists.}
\label{fig:DefEx}
\end{figure}

We show an example in \cref{fig:DefEx}. Taking $\alpha'=\frac{1}{2}$ and starting from the solution with $s=5/2$, we can perform a finite deformation by increasing or decreasing the squashing parameter. We keep a solution of the heterotic G$_2$ system if we deform the instantons in a very specific way as we change $s$. Note that deformation towards higher $s$ is unrestricted, whereas when decreasing $s$ we find that deformations beyond $s=\sqrt{\frac{3}{2}}$ are not allowed. therefore, the solutions with $s<1$ can not be accessed through a continuous deformation.

\subsection{Summary of results}\label{Summary}

We now summarize the results of this section, emphasizing the differences between the solutions obtained.

In \cref{subsec:cancan,subsec:canCO} we have described some isolated solutions where the string parameter $\alpha'$ can be made arbitrarily small by choosing a convenient representation of the gauge bundle. Nevertheless, from the point of view of heterotic string theory the gauge bundle representation must be contained in the adjoint representation of the gauge group $\text{E}_8\times\text{E}_8$. This imposes some restrictions on the representations we can actually choose, in particular in the dimension of the representation can not be arbitrarily large.

In \cref{subsec:canfam,subsec:COfam,subsec:famCO} we obtained solutions for a variety of ranges of squashing parameter $s$ and where the string parameter $\alpha'$ depends on the value of $s$. Some of the solutions are well-defined only in very small intervals for which the string parameter quickly blows up. These solutions are less interesting from a physical point of view, since $\alpha'$ should be understood as a small perturbation parameter.

For solutions defined on a wider range, it is interesting to analyze the behaviour in the limit where the squashing parameter becomes very large $s\rightarrow\infty$. Recall that this limit corresponds to the case where the $\text{SU}(2)\simeq\Sc^3$ fibres have large radius. Equivalently, the volume of the compact manifold blows up and the space decompactifies. In \cref{subsec:canfam,subsec:famCO} we find solutions where $\alpha'\rightarrow\infty$ in this limit. These solutions, although completely valid from a mathematical point of view, are not satisfactory for string theory as the parameter $\alpha'$ can not be made small.

In \cref{subsec:COfam} we find solutions where $\alpha'$ tends to a constant value as $s\rightarrow\infty$. The exact value of this constant depends on the gauge bundle representation and can therefore be chosen as small as desired. This gives raise to plenty of solutions where the string parameter is small. Furthermore, in \cref{subsec:canfam,subsec:COfam} we find solutions where $\alpha'\rightarrow 0$ as $s$ increases, providing further examples relevant for physics. These solutions correspond to a large volume limit: as the string parameter vanishes the volume of the compact manifold blows up.

It is also worth exploring the limit $s\rightarrow 0$. Geometrically, this corresponds to a situation where the $\text{SU}(2)\simeq\Sc^3$ fibres are shrunk to a point, thus having a singular space. For the solutions in \cref{subsec:canfam,subsec:COfam,subsec:famCO} where this limit can be considered, we always encounter $\alpha'\rightarrow 0$. Hence, the string parameter vanishes in the limit of maximum squashing.

At this point we would like to compare our results with the interesting solutions of Lotay and S\'a Earp \cite{Lotay:2021eog}. They construct solutions to the heterotic G$_2$ system on contact Calabi-Yau 7-folds, which are manifolds obtained from a circle fibration over a Calabi-Yau 3-fold. These present some common features with our solutions: first of all, whereas we rescale the metric along SU(2) fibres and our G$_2$ structure depends on the squashing parameter $s$, they perform a rescaling along the $U(1)$ fibre and their G$_2$ structure depends on the corresponding parameter---which they denote $\varepsilon$.

The authors of \cite{Lotay:2021eog} obtain connections which depend on the parameter $\varepsilon$, in the same way as our connections depend on the parameter $s$. However,  their tangent bundle connections satisfy the instanton condition only at first order in $\alpha'$, so they are approximate G$_2$-instantons. Our tangent bundle connections are exact G$_2$-instanton solutions satisfying the
first order in $\alpha'$ heterotic G$_2$ system given in \cref{sec:Heteroticg2Systems}.\footnote{In \cite{delaOssa:2014msa} it was argued that these equations remain valid to all orders in $\alpha'$ after suitable field redefinitions.}

In \cite{Lotay:2021eog}, the authors find solutions with an AdS$_3$ spacetime where the string parameter $\alpha'(\varepsilon)$ depends on the parameter $\varepsilon$, as is the case for most of our solutions where $\alpha'(s)$ depends on $s$. The behaviour $\alpha'\rightarrow 0$ that we find when $s\rightarrow 0$ is also present in their solutions when $\varepsilon\rightarrow 0$. Nevertheless, the curvature of the AdS$_3$ tends to zero in their solutions whereas it blows up in our case. As we have mentioned before, our solutions are such that that spacetime can never be  Minkowski.

Finally, let us stress that in all the solutions of \cite{Lotay:2021eog} there is a mutual dependence between the parameters $\alpha'$ and $\varepsilon$. Our solutions in \cref{subsec:canfam,subsec:COfam,subsec:famCO} present an analogous feature in terms of $\alpha'$ and $s$. We have also managed to construct solutions where $\alpha'$ does not actually depend on the squashing parameter $s$.

This is the case of the solutions of \cref{subsec:famfam}, which deserve special attention. By using two different instanton connections from the one-parameter family, we construct a family of solutions where $s$ and $\alpha'$ are two parameters that can be chosen independently within certain region. This provides a large family of solutions. In addition, we can regard these backgrounds as finite deformations: take any solution as the starting point. We can then keep the string parameter $\alpha'$ fixed while deforming the G$_2$ structure by modifying the squashing parameter $s$. This also results in a deformation of the instanton connections.

These are the first explicit examples of families of finite deformations (together with a very particular solution described in \cref{subsec:COfam}) and they provide a finite version of the infinitesimal deformations discussed in \cite{delaOssa:2017pqy}. Making a thorough study of the relation between infinitesimal and finite deformations will be the subject of future work.

\bigskip

\section{Conclusions and outlook}
\label{sec:conclusions}

In this paper we have constructed new solutions of the heterotic G$_2$ system on squashed homogeneous 3-Sasakian manifolds. We have given a detailed description of the specific family of G$_2$ structures used in the construction as well as the instanton connections employed. These solutions are grouped into different families in terms of the choice of instanton connections in the gauge and tangent bundles. The resulting spacetime is AdS$_3$ for all the solutions found and, together with those of \cite{Lotay:2021eog}, constitute the only few explicit examples of this type.

Moreover, these families of solutions include the first examples of heterotic G$_2$ backgrounds that can be described as finite deformations from a given solution. For any particular solution of those presented in \cref{subsec:famfam}---as well as for the solution depicted in \cref{fig:oneparCOc8} in \cref{subsec:COfam}---we can keep the string parameter $\alpha'$ fixed and obtain new, deformed solutions by performing a squashing of the metric and deforming the bundle instantons.

In \cite{delaOssa:2021cgd} it is explained how heterotic G$_2$ solutions can secretely present enhanced $\mathcal{N}=2$ supersymmetry. The authors showed that this was indeed the case for one of the solutions constructed in \cite{Fernandez:2008wla}. Nevertheless, the supersymmetry enhancement requires the 3-dimensional spacetime to be Minkowski. This is not the case for our solutions, so we can guarantee these are proper $\mathcal{N}=1$ background solutions of the heterotic G$_2$ system.

There are several possible future directions to pursuit. First of all, it would be very interesting to use our solutions to improve our understanding of the moduli space of heterotic G$_2$ systems. An easy first step would be to compute the infinitesimal version of the deformation described by our solutions and relate it to the formalism of \cite{delaOssa:2017pqy}. One could also try to explicitly compute the cohomologies encoding the infinitesimal deformations.

We have obtained solutions that follow one particular deformation direction, but we expect further deformations to be allowed. A possibility would be to look into alternative deformations of the G$_2$ structure, such as the ones described for the Aloff-Wallach space in \cite{Ball:2016xte}, and try to find solutions extending the ones we have constructed.

Alternatively, we could explore further choices of instanton connections and study if these provide new solutions of the heterotic G$_2$ system. One possibility would be to use the instantons in the tangent bundle we described at the end of \cref{sec:tangentbundleinstantons}. Further options are available in the literature: for example a 15-dimensional family of G$_2$-instantons for the 7-sphere is described in \cite{2020arXiv200202386W}. Additionally, a detailed description of infinitesimal deformations of G$_2$-instantons for nearly-parallel G$_2$-manifolds can be found \cite{singhal2021deformations}. Finally, connections on homogeneous 3-Sasakian manifolds are described in some generality in \cite{2018arXiv180110526D}.

A natural direction---beyond the study of deformations---would be to generalize our construction to obtain completely different solutions. A first option would be to consider alternative G$_2$ structures (not necessarily connected to our family via deformations) such as the standard nearly parallel G$_2$ structures of \cite{2010JGP....60..326A}. Going further, one could try to extend our construction to squashed non-homogeneous 3-Sasakian manifolds, or to generalized 3-($\alpha$,$\delta$) Sasakian manifolds: \cite{2018arXiv180406700A,2019arXiv190307815D}. It would also be interesting to construct solutions involving Sasaki or Sasaki-Einstein manifolds, or more general Aloff-Wallach spaces $N_{p,q}$ making use of the connections of \cite{Ball:2016xte}. Another option would be to generalize the solutions of \cite{Fernandez:2008wla, Fernandez:2014pfa} to other nilmanifolds using the detailed study of \cite{2020arXiv200615925D}.  Finally, one could attempt to exploit the existence of almost contact metric 3-structures on manifolds with a G$_2$-structure to try to generalize the example of section 5.1 of \cite{delaOssa:2021cgd}.

In \cite{Clarke:2020erl} T-dual solutions of the heterotic G$_2$ system are constructed. Applying a T-duality to our families of solutions---for example on the circle inside the 3-Sasakian fibre---to obtain new T-dual solutions is a tantalizing possibility. One could also attempt to perform a full non-abelian T-duality on the SU(2) fibre \cite{delaOssa:1992vci}.

Finally, compactifications on manifolds with a $G$-structure present additional worldsheet symmetries \cite{Howe:1991ic,delaOssa:2018azc}. In the case of $G$-holonomy, it has been argued that this corresponds to extensions of the worldsheet superconformal algebra by additional chiral operators \cite{Odake:1988bh,Shatashvili:1994zw,Figueroa-OFarrill:1996tnk}. It would be interesting to perform a worldsheet study of the solutions we have presented. This question has been addressed for G$_2$ twisted connected sums in \cite{Fiset:2018huv}, and for G$_2$ extra twisted connected sums as well as Spin(7) generalized connected sum manifolds in \cite{Fiset:2021ruv}. The AdS$_3$ spacetime factor would require the presence of the superconformal algebra introduced in \cite{Fiset:2021azq}.

\bigskip

\section*{Acknowledgements}

We would like to thank Chris Beem, Philip Candelas, Benoit Charbonneau, Marc-Antoine Fiset and Gon\c{c}alo Oliveira for discussions. MG is supported by a scholarship from the Mathematical Institute, University of Oxford, as well as a fellowship from ``la Caixa” Foundation (ID 100010434) with fellowship code LCF/BQ/EU17/11590062.

\appendix

\bigskip

\section{Sp(2) structure equations}
\label{sec:Sp2struceq}

A local coframe for Sp(2) is obtained from a basis of the Lie algebra $\mathfrak{sp}(2)$ as described in \cref{sec:Homogeneouscase}. We choose the following basis
\begin{align*}
    e^1&=\begin{pmatrix}
i &0 \\
0 &0
\end{pmatrix}, \qquad 
&e^2=\begin{pmatrix}
j &0 \\
0 &0
\end{pmatrix}, \qquad 
e^3&=\begin{pmatrix}
k &0 \\
0 &0
\end{pmatrix}, \\
e^4&=\begin{pmatrix}
0 &-1 \\
1 &0
\end{pmatrix}, \qquad 
&e^5=\begin{pmatrix}
0 &i \\
i &0
\end{pmatrix}, \qquad 
e^6&=\begin{pmatrix}
0 &j \\
j &0
\end{pmatrix}, \qquad 
e^7=\begin{pmatrix}
0 &-k \\
-k &0
\end{pmatrix}, \\
e^8&=\begin{pmatrix}
0 &0 \\
0 &i
\end{pmatrix}, \qquad 
&e^9=\begin{pmatrix}
0 &0 \\
0 &j
\end{pmatrix}, \qquad 
e^{10}&=\begin{pmatrix}
0 &0 \\
0 &k
\end{pmatrix}.
\end{align*}

The nonzero structure constants on this basis are 
\begin{align}
\label{eq:structureconstantssp2}
\begin{split}
2&=f_{23}^1=-f_{45}^1=-f_{67}^1=f_{31}^2=-f_{46}^2=f_{57}^2=f_{12}^3=f_{47}^3=f_{56}^3 \, ,\\
1&=f_{15}^4=f_{26}^4=-f_{37}^4=-f_{14}^5=-f_{27}^5=-f_{36}^5=f_{17}^6=-f_{24}^6=f_{35}^6=-f_{16}^7=f_{25}^7=f_{34}^7 \, ,\\
1&=f_{58}^4=f_{69}^4=-f_{7 \ 10}^4=-f_{48}^5=f_{6 \ 10}^5=f_{79}^5=-f_{49}^6=-f_{5 \ 10}^6=-f_{78}^6=f_{4 \ 10}^7=-f_{59}^7=f_{68}^7 \, ,\\
2&=f_{9 \ 10}^8=f_{45}^8=-f_{6 7}^8=f_{10 \ 8}^9=f_{4  6}^{9}=f_{5 7}^{9}=f_{8 9}^{10}=-f_{4  7}^{10}=f_{56}^{10} \, .
\end{split}
\end{align}
Substituting \eqref{eq:structureconstantssp2} in \eqref{eq:generalstructureequation} we obtain the structure equations of the Sp(2) coframe\footnote{Note that the normalization of our basis, convenient for the description of the 3-Sasakian structure, makes our choice of metric not proportional to the Killing form of Sp(2).}
\begin{align}
\label{sp2structureeqs}
\begin{split}
\dd e^1&=-2 \, e^2\wedge e^3+2 \, e^4\wedge e^5+2 \, e^6\wedge e^7, \\
\dd e^2&=-2 \, e^3\wedge e^1+2 \, e^4\wedge e^6-2 \, e^5\wedge e^7, \\
\dd e^3&=-2 \, e^1\wedge e^2-2 \, e^4\wedge e^7-2 \, e^5\wedge e^6, \\
\dd e^4&=-e^1\wedge e^5-e^2\wedge e^6+e^3\wedge e^7-e^5\wedge e^8-e^6\wedge e^9+e^7\wedge e^{10}, \\
\dd e^5&=+e^1\wedge e^4+e^2\wedge e^7+e^3\wedge e^6+e^4\wedge e^8-e^7\wedge e^9-e^6\wedge e^{10}, \\
\dd e^6&=-e^1\wedge e^7+e^2\wedge e^4-e^3\wedge e^5+e^7\wedge e^8+e^4\wedge e^9+e^5\wedge e^{10}, \\
\dd e^7&=+e^1\wedge e^6-e^2\wedge e^5-e^3\wedge e^4-e^6\wedge e^8+e^5\wedge e^9-e^4\wedge e^{10}, \\
\dd e^8&=-2 \, e^9\wedge e^{10}-2 \, e^4\wedge e^5+2 \, e^6\wedge e^7, \\
\dd e^9&=-2 \, e^{10}\wedge e^8-2 \, e^4\wedge e^6-2 \, e^5\wedge e^7, \\
\dd e^{10}&=-2 \, e^8\wedge e^9+2 \, e^4\wedge e^7-2 \, e^5\wedge e^6.
\end{split}
\end{align}

\bigskip

\section{SU(3) structure equations}
\label{SU3struceq}

We construct a coframe for SU(3) using a convenient basis of the Lie algebra $\mathfrak{su}(3)$, given by an appropriate normalization of the Gell-mann matrices
\begin{align*}
e^1&=\begin{pmatrix}
i &0 &0 \\
0 &-i &0 \\
0 &0 &0
\end{pmatrix}, &
e^2&=\begin{pmatrix}
0 &1 &0 \\
-1 &0 &0 \\
0 &0 &0
\end{pmatrix},  &
e^3&=\begin{pmatrix}
0 &i &0 \\
i &0 &0 \\
0 &0 &0
\end{pmatrix},\\
e^4&=\sqrt{2}\begin{pmatrix}
0 &0 &i \\
0 &0 &0 \\
i &0 &0
\end{pmatrix}, &
e^5&=\sqrt{2}\begin{pmatrix}
0 &0 &1 \\
0 &0 &0 \\
-1 &0 &0
\end{pmatrix}, &
e^6&=\sqrt{2}\begin{pmatrix}
0 &0 &0 \\
0 &0 &i \\
0 &i &0
\end{pmatrix},\\
e^7&=\sqrt{2}\begin{pmatrix}
0 &0 &0 \\
0 &0 &-1 \\
0 &1 &0
\end{pmatrix},  &
e^8&=-\frac{i}{3}\begin{pmatrix}
1 &0 &0 \\
0 &1 &0 \\
0 &0 &-2
\end{pmatrix}. & &
\end{align*}
The nonzero structure constants satisfy $f_{\mu\nu}^\rho=-f_{\nu\mu}^\rho$ and take the values:
\begin{align}
\label{eq:structureconstantssu3}
\begin{split}
2&=f_{23}^1=-f_{45}^1=-f_{67}^1=f_{31}^2=-f_{46}^2=f_{57}^2=f_{12}^3=f_{47}^3=f_{56}^3 \, ,\\
1&=f_{15}^4=f_{26}^4=-f_{37}^4=-f_{14}^5=-f_{27}^5=-f_{36}^5=f_{17}^6=-f_{24}^6=f_{35}^6=-f_{16}^7=f_{25}^7=f_{34}^7 \, ,\\
1&=f_{58}^4=-f_{48}^5=-f_{78}^6=f_{68}^7 \, ,\\
6&=f_{45}^8=-f_{6 7}^8.
\end{split}
\end{align}
Now from \eqref{eq:generalstructureequation} we deduce that the corresponding basis of left-invariant one-forms satisfy the following structure equations:
\begin{align}
\label{su3structureeqs}
\begin{split}
\dd e^1&=-2 \, e^2\wedge e^3+2 \, e^4\wedge e^5+2 \, e^6\wedge e^7 \, , \\
\dd e^2&=-2 \, e^3\wedge e^1+2 \, e^4\wedge e^6-2 \, e^5\wedge e^7 \, , \\
\dd e^3&=-2 \, e^1\wedge e^2-2 \, e^4\wedge e^7-2 \, e^5\wedge e^6 \, , \\
\dd e^4&=-e^1\wedge e^5-e^2\wedge e^6+e^3\wedge e^7- e^5\wedge e^8 \, , \\
\dd e^5&=+e^1\wedge e^4+e^2\wedge e^7+e^3\wedge e^6+ e^4\wedge e^8 \, , \\
\dd e^6&=-e^1\wedge e^7+e^2\wedge e^4-e^3\wedge e^5+ e^7\wedge e^8 \, , \\
\dd e^7&=+e^1\wedge e^6-e^2\wedge e^5-e^3\wedge e^4- e^6\wedge e^8 \, , \\
\dd e^8&=-6 \,  e^4\wedge e^5+6 \,  e^6\wedge e^7 \, .
\end{split}
\end{align}

\newpage

\section{Explicit representation matrices of bundle adjoint action}

\subsection{Canonical connection}
\label{sec:ExpRepMatSp1AdjAct}

From the structure constants of the Lie algebra of Sp(2) \eqref{eq:structureconstantssp2} we can read off the adjoint action of SU(2) on the tangent bundle of the squashed 7-sphere, given by the following matrices
{\small
\begin{equation}
I_8=\begin{pmatrix}
0 &0 &0 &0 &0 &0 &0 \\
0 &0 &0 &0 &0 &0 &0 \\
0 &0 &0 &0 &0 &0 &0 \\
0 &0 &0 &0 &-1 &0 &0 \\
0 &0 &0 &1 &0 &0 &0 \\
0 &0 &0 &0 &0 &0 &1 \\
0 &0 &0 &0 &0 &-1 &0 \\
\end{pmatrix}, \quad
I_9=\begin{pmatrix}
0 &0 &0 &0 &0 &0 &0 \\
0 &0 &0 &0 &0 &0 &0 \\
0 &0 &0 &0 &0 &0 &0 \\
0 &0 &0 &0 &0 &-1 &0 \\
0 &0 &0 &0 &0 &0 &-1 \\
0 &0 &0 &1 &0 &0 &0 \\
0 &0 &0 &0 &1 &0 &0 \\
\end{pmatrix}, \quad
I_{10}=\begin{pmatrix}
0 &0 &0 &0 &0 &0 &0 \\
0 &0 &0 &0 &0 &0 &0 \\
0 &0 &0 &0 &0 &0 &0 \\
0 &0 &0 &0 &0 &0 &1 \\
0 &0 &0 &0 &0 &-1 &0 \\
0 &0 &0 &0 &1 &0 &0 \\
0 &0 &0 &-1 &0 &0 &0 \\
\end{pmatrix}.
\end{equation}
}
It can be checked that these matrices satisfy the normalization $[I_a, I_b]=2\,\epsilon\indices{_a_b^c}\,I_c$, and we also find
\begin{equation}
\tr(I_a I_b)=-4\,\delta_{ab}\,.
\end{equation}

For the squashed Aloff-Wallach space case, we use the structure constants \eqref{eq:structureconstantssu3} and find the action described by
{\small
\begin{equation}
I_8=\begin{pmatrix}
0 &0 &0 &0 &0 &0 &0 \\
0 &0 &0 &0 &0 &0 &0 \\
0 &0 &0 &0 &0 &0 &0 \\
0 &0 &0 &0 &-1 &0 &0 \\
0 &0 &0 &1 &0 &0 &0 \\
0 &0 &0 &0 &0 &0 &1 \\
0 &0 &0 &0 &0 &-1 &0 \\
\end{pmatrix},
\end{equation}
}
and we have
\begin{equation}
\tr(I_8 I_8)=-4\,.
\end{equation}

\subsection{Clarke-Oliveira connection}
\label{sec:ExpRepMatCOconn}

From the structure constants of the Lie algebra of Sp(2) \eqref{eq:structureconstantssp2} and SU(3) \eqref{eq:structureconstantssu3} we can read off the adjoint action on the tangent bundle of the SU(2) associated to the 3-Sasakian triple of Killing fields. These matrices are valid for both the squashed 7-sphere and the squashed Aloff-Wallach space
{\small
\begin{equation}
I_1=\begin{pmatrix}
0 &0 &0 &0 &0 &0 &0 \\
0 &0 &-2 &0 &0 &0 &0 \\
0 &2 &0 &0 &0 &0 &0 \\
0 &0 &0 &0 &1 &0 &0 \\
0 &0 &0 &-1 &0 &0 &0 \\
0 &0 &0 &0 &0 &0 &1 \\
0 &0 &0 &0 &0 &-1 &0 \\
\end{pmatrix}, ~ ~ ~
I_2=\begin{pmatrix}
0 &0 &2 &0 &0 &0 &0 \\
0 &0 &0 &0 &0 &0 &0 \\
-2 &0 &0 &0 &0 &0 &0 \\
0 &0 &0 &0 &0 &1 &0 \\
0 &0 &0 &0 &0 &0 &-1 \\
0 &0 &0 &-1 &0 &0 &0 \\
0 &0 &0 &0 &1 &0 &0 \\
\end{pmatrix}, ~ ~ ~
I_3=\begin{pmatrix}
0 &-2 &0 &0 &0 &0 &0 \\
2 &0 &0 &0 &0 &0 &0 \\
0 &0 &0 &0 &0 &0 &0 \\
0 &0 &0 &0 &0 &0 &-1 \\
0 &0 &0 &0 &0 &-1 &0 \\
0 &0 &0 &0 &1 &0 &0 \\
0 &0 &0 &1 &0 &0 &0 \\
\end{pmatrix}.
\end{equation}
}
It can be checked that these matrices satisfy the normalization $[I_i, I_j]=2\,\epsilon\indices{_i_j^k}\,I_k$, and we also find
\begin{equation}
\tr(I_i I_j)=-12\,\delta_{ij}\,.
\end{equation}

\bigskip

\section{Most general G${}_2$-compatible metric connection}
\label{sec:ConnectionCurvatureg2compatible}

We introduced in \cref{sec:tangentbundleinstantons} a 1-parameter family of instantons in the tangent bundle given by metric connections compatible with the G${}_2$-structure. We list here their explicit expression, obtained from substituting the Levi-Civita connection and contorsion in \eqref{eq:oneformfinalformula}. Let
\begin{equation}
    \kappa(a,s)=(1+10\,a)s+(1-2\,a)\frac{1}{s}\,,
\end{equation}
the nonzero connection one-forms (up to antisymmetry $\omega\indices{^\mu_\nu}=-\omega\indices{^\nu_\mu}$) for the squashed 7-sphere are given by
\begin{align*}
\omega\indices{^1_2}&=-\kappa(a,s)\,\eta^3\,,  & \omega\indices{^2_3}&=-\kappa(a,s)\,\eta^1\,, \\
\omega\indices{^3_1}&=-\kappa(a,s)\,\eta^2\,,  & & \\
\omega\indices{^4_5}&=\frac{1}{2}\kappa(a,s)\,\eta^1 -\eta^8\,, &  \omega\indices{^6_7}&=\frac{1}{2}\kappa(a,s)\,\eta^1 +\eta^8\,, \\
\omega\indices{^4_6}&=\frac{1}{2}\kappa(a,s) \, \eta^2 -\eta^9\,, & \omega\indices{^5_7}&=-\frac{1}{2}\kappa(a,s)\,\eta^2 -\eta^9\,, \\
\omega\indices{^4_7}&=-\frac{1}{2}\kappa(a,s) \, \eta^3 +\eta^{10} \, , & \omega\indices{^5_6}&=-\frac{1}{2}\kappa(a,s) \, \eta^3 -\eta^{10} \, ,
\end{align*}
and for the squashed Aloff-Wallach space they are given by
\begin{align*}
\omega\indices{^1_2}&=-\kappa(a,s) \, \eta^3 \, ,  & \omega\indices{^2_3}&=-\kappa(a,s) \, \eta^1 \, , \\
\omega\indices{^3_1}&=-\kappa(a,s) \, \eta^2 \, ,  & & \\
\omega\indices{^4_5}&=\frac{1}{2}\kappa(a,s) \, \eta^1 -\eta^8 \, , &  \omega\indices{^6_7}&=\frac{1}{2}\kappa(a,s) \, \eta^1 +\eta^8 \, , \\
\omega\indices{^4_6}&=\frac{1}{2}\kappa(a,s) \, \eta^2 \, , & \omega\indices{^5_7}&=-\frac{1}{2}\kappa(a,s) \, \eta^2 \, , \\
\omega\indices{^4_7}&=-\frac{1}{2}\kappa(a,s) \, \eta^3 \, , & \omega\indices{^5_6}&=-\frac{1}{2}\kappa(a,s) \, \eta^3 \, ,
\end{align*}

From $F=\dd \omega+\omega\wedge\omega$ we compute the curvature. For the squashed 7-sphere (up to antisymmetry $F_{\mu\nu}=-F_{\nu\mu}$) the nonzero terms are
\begin{align*}
F_{12}&=-\kappa(a,s)\left[\left(\kappa(a,s)-\frac{2}{s}\right)\eta^1\wedge\eta^2+2 \, s \, \omega^3\right] \, , \\
F_{23}&=-\kappa(a,s)\left[\left(\kappa(a,s)-\frac{2}{s}\right)\eta^2\wedge\eta^3+2 \, s \, \omega^1\right] \, , \\
F_{31}&=-\kappa(a,s)\left[\left(\kappa(a,s)-\frac{2}{s}\right)\eta^3\wedge\eta^1+2 \, s \, \omega^2\right] \, , \\
F_{45}&=\frac{1}{2}\kappa(a,s)\left(\kappa(a,s)-\frac{2}{s}\right)\eta^2\wedge\eta^3 +s\left(\kappa(a,s)+\frac{2}{s}\right)\eta^4\wedge\eta^5 +s\left(\kappa(a,s)-\frac{2}{s}\right)\eta^6\wedge\eta^7 \, , \\
F_{67}&=\frac{1}{2}\kappa(a,s)\left(\kappa(a,s)-\frac{2}{s}\right)\eta^2\wedge\eta^3 +s\left(\kappa(a,s)-\frac{2}{s}\right)\eta^4\wedge\eta^5 +s\left(\kappa(a,s)+\frac{2}{s}\right)\eta^6\wedge\eta^7 \, , \\
F_{46}&=\frac{1}{2}\kappa(a,s)\left(\kappa(a,s)-\frac{2}{s}\right)\eta^3\wedge\eta^1 +s\left(\kappa(a,s)+\frac{2}{s}\right)\eta^4\wedge\eta^6 -s\left(\kappa(a,s)-\frac{2}{s}\right)\eta^5\wedge\eta^7 \, , 
\end{align*}

\begin{align*}
F_{57}&=-\frac{1}{2}\kappa(a,s)\left(\kappa(a,s)-\frac{2}{s}\right)\eta^3\wedge\eta^1 -s\left(\kappa(a,s)-\frac{2}{s}\right)\eta^4\wedge\eta^6 +s\left(\kappa(a,s)+\frac{2}{s}\right)\eta^5\wedge\eta^7 \, , \\
F_{47}&=-\frac{1}{2}\kappa(a,s)\left(\kappa(a,s)-\frac{2}{s}\right)\eta^1\wedge\eta^2 +s\left(\kappa(a,s)+\frac{2}{s}\right)\eta^4\wedge\eta^7 +s\left(\kappa(a,s)-\frac{2}{s}\right)\eta^5\wedge\eta^6 \, , \\
F_{56}&=-\frac{1}{2}\kappa(a,s)\left(\kappa(a,s)-\frac{2}{s}\right)\eta^1\wedge\eta^2 +s\left(\kappa(a,s)-\frac{2}{s}\right)\eta^4\wedge\eta^7 +s\left(\kappa(a,s)+\frac{2}{s}\right)\eta^5\wedge\eta^6 \, ,
\end{align*}
whereas for the squashed Aloff-Wallach space we find
\begin{align*}
F_{12}&=-\kappa(a,s)\left[\left(\kappa(a,s)-\frac{2}{s}\right)\eta^1\wedge\eta^2+2 \, s \, \omega^3\right] \, , \\
F_{23}&=-\kappa(a,s)\left[\left(\kappa(a,s)-\frac{2}{s}\right)\eta^2\wedge\eta^3+2 \, s \, \omega^1\right] \, , \\
F_{31}&=-\kappa(a,s)\left[\left(\kappa(a,s)-\frac{2}{s}\right)\eta^3\wedge\eta^1+2 \, s \, \omega^2\right] \, , \\
F_{45}&=\frac{1}{2}\kappa(a,s)\left(\kappa(a,s)-\frac{2}{s}\right)\eta^2\wedge\eta^3 +s\left(\kappa(a,s)+\frac{6}{s}\right)\eta^4\wedge\eta^5 +s\left(\kappa(a,s)-\frac{6}{s}\right)\eta^6\wedge\eta^7 \, , \\
F_{67}&=\frac{1}{2}\kappa(a,s)\left(\kappa(a,s)-\frac{2}{s}\right)\eta^2\wedge\eta^3 +s\left(\kappa(a,s)-\frac{6}{s}\right)\eta^4\wedge\eta^5 +s\left(\kappa(a,s)+\frac{6}{s}\right)\eta^6\wedge\eta^7 \, , \\
F_{46}&=\frac{1}{2}\kappa(a,s)\left[\left(\kappa(a,s)-\frac{2}{s}\right)\eta^3\wedge\eta^1+2 \, s \, \omega^2\right] \, , \\
F_{57}&=-\frac{1}{2}\kappa(a,s)\left[\left(\kappa(a,s)-\frac{2}{s}\right)\eta^3\wedge\eta^1+2 \, s \, \omega^2\right] \, , \\
F_{47}&=-\frac{1}{2}\kappa(a,s)\left[\left(\kappa(a,s)-\frac{2}{s}\right)\eta^1\wedge\eta^2+2 \, s \, \omega^3\right] \, , \\
F_{56}&=-\frac{1}{2}\kappa(a,s)\left[\left(\kappa(a,s)-\frac{2}{s}\right)\eta^1\wedge\eta^2+2 \, s \, \omega^3\right] \, .
\end{align*}

\newpage

\section{Coefficients of cubic equations}
\label{sec:cubiceqs}

Here we list the coefficients of the cubic equations that need to be satisfied in order to solve the heterotic Bianchi identity in \cref{subsec:canfam,subsec:COfam,subsec:famCO}.

The coefficients of the positive powers of $a$ are the same for all the cubics.

\begin{table}[h]
{
\centering
\renewcommand{\arraystretch}{2.2}
\begin{tabular}{|c|c|}
\hline 
Power of $a$ &
Coefficient \\
\hline
\hline
$a^3$ & $\dfrac{12 \left(5 \,  s^2-1\right)^3}{s^2 \left(2 \,  s^2-1\right)}$ \\ \hline
$a^2$ & $\dfrac{6 \left(1-5 s^2\right)^2 \left(7
   s^2-1\right)}{s^2 \left(2 \,  s^2-1\right)}$ \\ \hline
$a$ & $\dfrac{3 \left(55 \,  s^6+19 \,  s^4-31
    \, s^2+5\right)}{s^2 \left(2 \,  s^2-1\right)}$  \\
\hline
\end{tabular}
\renewcommand{\arraystretch}{1}
\caption{List of cubic coefficients for positive powers of $a$.
}
\label{tab:cubicpositivecoeffs}
}
\end{table}

\begin{table}[h]
{
\centering
\renewcommand{\arraystretch}{2.2}
\begin{tabular}{|c|c|c|}
\hline 
Section &
Independent term &
Coefficient \\
\hline
\hline
\multirow{2}{*}{\ref{subsec:canfam}} 
  & 7-sphere & $\dfrac{-4 \,  c \,  s^2+2 c-15 \,  s^6-21 \,  s^4+19 \,  s^2+1}{2 \,  s^2-4 s^4}$ \\ \cline{2-3}
  & Aloff-Wallach space & $\dfrac{3 \left(q \left(4 \,  s^2-2\right)+5 \,  s^6+7 \,  s^4-17 \,  s^2+5\right)}{4 \,  s^4-2 \,  s^2}$ \\
\hline
\multirow{2}{*}{\ref{subsec:COfam}} 
  & 7-sphere & $\dfrac{4 \,  c \,  s^2-2 c-15 \,  s^6-21 \,  s^4+19 \,  s^2+1}{2 \,  s^2-4 \,  s^4}$ \\ \cline{2-3}
  & Aloff-Wallach space & $\dfrac{q \left(2-4 \,  s^2\right)+3 \left(5 \,  s^6+7 \,  s^4-17 \,  s^2+5\right)}{4 \,  s^4-2 \,  s^2}$ \\[1ex]
\hline
\multirow{2}{*}{\ref{subsec:famCO}} 
  & 7-sphere & $\dfrac{15 \,  s^6+117 \,  s^4-115 \,  s^2+23}{4 \,  s^4-2 \,  s^2}$ \\ \cline{2-3}
  & Aloff-Wallach space & $\dfrac{3 \left(5 \,  s^6+39 \,  s^4-49 \,  s^2+13\right)}{4 \,  s^4-2 \,  s^2}$ \\
\hline
\end{tabular}
\renewcommand{\arraystretch}{1}
\caption{List of independent terms of the cubics.
}
\label{tab:cubiccoeffs}
}
\end{table}

\newpage

\providecommand{\href}[2]{#2}\begingroup\raggedright\endgroup

\end{document}